\documentclass[letterpaper,11pt]{article}

\usepackage{amsmath}
\usepackage{amssymb}
\usepackage[vmargin=1in,hmargin=1in]{geometry}
\usepackage{graphicx}
\usepackage{color}
\usepackage{url}
\usepackage{natbib}
\usepackage{soul}
\usepackage{bibtopic}

\usepackage{algorithm,algorithmic}
\makeatletter
\newcommand{\FORCONC}[2][default]{
  \ALC@it\algorithmicfor\ #2\ %
  \algorithmicdo\ \textbf{concurrently}%
  \ALC@com{#1}\begin{ALC@for}}
\makeatother
  
\usepackage{amsthm}

\newtheorem{lemma}{Lemma}
\newtheorem{theorem}{Theorem}

\newtheorem{remark}{Remark}

\newcommand{\xbf}{{\mathbf{x}}}
\newcommand{\ybf}{{\mathbf{y}}}
\DeclareMathOperator*{\argmin}{argmin} 
\DeclareMathOperator*{\argmax}{argmax} 

\allowdisplaybreaks

\usepackage{fancyhdr}
\begin{document}

\author{Evangelos Evangelou$^{1}$
  \hspace{1pt}
  and Vasileios Maroulas$^{2}$\\[5pt]
$^1$ \small{Department of Mathematical Sciences, University of Bath, Bath BA2 7AY, UK.}\\
$^2$ \small{Department of Mathematics, University of Tennessee, Knoxville, TN 37996, USA.}}
\title{Sequential Empirical Bayes Method for Filtering Dynamic Spatiotemporal Processes}
\date{15 June 2017}

\footnotetext{\textit{Address for correspondence}: Evangelos Evangelou,
  Department of Mathematical Sciences, University of Bath, Bath BA2 7AY,
  UK. \texttt{email:}~\url{ee224@bath.ac.uk}}

\maketitle

\vfill

\noindent\textbf{Acknowledgements:} This research was conducted during the
second author's visit as a 
Leverhulme Trust Visiting Fellow at the Department of Mathematical Sciences
at the University of Bath whose hospitality is greatly appreciated. Both
authors would like to thank the Leverhulme Trust for partial financial
support, Grant \# VF-2012-006. The second author would like to also thank
the Simons Foundation for partial financial support with Grant \# 279870,
and the Air Force Office of Scientific Research for partial financial
support with Grant \# FA9550-15-1-0103.

\clearpage

\begin{abstract}
  We consider online prediction of a latent dynamic spatiotemporal process
  and estimation of the associated model parameters based on noisy data.
  The problem is motivated by the analysis of spatial data arriving in
  real-time and the current parameter estimates and predictions are updated
  using the new data at a fixed computational cost. Estimation and
  prediction is performed within an empirical Bayes framework with the aid
  of Markov chain Monte Carlo samples. Samples for the latent spatial field
  are generated using a sampling importance resampling algorithm with a
  skewed-normal proposal and for the temporal parameters using Gibbs
  sampling with their full conditionals written in terms of sufficient
  quantities which are updated online. The spatial range parameter is
  estimated by a novel online implementation of an empirical Bayes method,
  called herein \emph{sequential empirical Bayes} method. A simulation
  study shows that our method gives similar results as an offline Bayesian
  method. We also find that the skewed-normal proposal improves over the
  traditional Gaussian proposal. The application of our method is
  demonstrated for online monitoring of radiation after the Fukushima
  nuclear accident.
\end{abstract}

\begin{quote}
\end{quote}


\section{Introduction}\label{intro}

Many problems related to ecology, epidemiology, defense, and economics
exhibit a simultaneous variability in space and time \citep{CrWi,ShZkbook}.
Daily levels of precipitation, temperature, or other environmental
variables across a region in a year, e.g. the monitoring of pollutants, the
estimation of trajectories of biological entities, or the monitoring of
mobile threats within a sensor network \citep{Paci201379,RMS,MaNe,RMS2} are a
few of a gamut of paradigms which require the careful treatment of
spatiotemporal processes in real time. 

Our motivation for this study is the monitoring of radiation after the
Fukushima nuclear accident. Following the accident, radioactive material
was released in the environment which rendered the surrounding areas
inhabitable. The authorities established a monitoring program to measure
radiation by sampling at specific locations within the infected area. These
samples, which are collected daily, can be used to produce a heatmap
indicating dangerous zones. This requires spatiotemporal modeling of
radiation in real-time to incorporate the new information as soon as it is
received. As radiation is measured by the number of nuclear decays emitted
each second, a Gaussian model would be inappropriate in this case and a
more flexible model is needed.

In cases concerning the study of radiation, disease spreading, weather, and
target tracking, where there is no indication when the study will terminate
while real-time information must be incorporated to facilitate a
rapid-response system, online methods which can assimilate the new data in
a fixed computational cost are urgently needed over offline methods.
Indeed, the computing costs of offline methods increase commensurately with
time and become slower as more and more data are collected. Especially in
the modern days of big data, offline methods become inefficient to run for
long time periods while online methods are able to take advantage the
often-assumed Markov property of the model to simplify computations.

The general setup consists of a latent dynamic process with Markovian
evolution while the observation process possesses a conditional
independence property. This setup is known as state-space or hidden Markov
model. The problem is to compute the filtering distribution, i.e. the
distribution of the current value of the latent process given the full
observation history up to the present time. A well-known example of an
online method is the Kalman filter which gives the exact filtering
distribution in the case of linear Gaussian models. A widely-applicable
alternative is particle filtering, or sequential Monte-Carlo, which is a
sequential importance sampling method that approximates the filtering
distribution by a set of weighted samples \citep{DFG}. On the other hand,
it is well known that particle filtering does not perform well when the
dimension of the latent process is large, as is the case of many
spatiotemporal applications \citep[Section~8.4.6]{CrWi}.

In addition to the filtering problem, it is also required to estimate the
unknown parameters of the model using the observed data on the fly. In a
Bayesian setting, this amounts to computing the posterior distribution of
these parameters, however the problem now becomes significantly more
difficult as there is no straightforward way of integrating out the latent
process in an online fashion. 
A comprehensive review of online estimation methods can be found in
\cite{K14}. An obvious solution is to augment the latent process with the
parameters, and estimate the joint distribution together by standard
particle methods. However, this was recognized by \cite{Kit} that this
procedure leads to erroneous estimation.
An alternative method, proposed by \cite{liuwest}, is to impose
artificial dynamics on the parameters and estimate them simultaneously with
the latent process using particle filters. However, this solution
introduces bias in the estimation and requires a significant amount of
tuning. The alternative, proposed by \cite{Svo, Fea}, is to write the full
conditionals of the parameters in terms of sufficient statistics which can
be updated sequentially. This approach facilitates sampling from the full
conditional of the parameters, however it does not avoid the degeneracy
problem and is only applicable to those models where sufficient statistics
exist, which is not the case for spatial models where the spatial range
parameter is unknown.

This manuscript focuses on the methodology for \emph{online} filtering of a
dynamic spatiotemporal process and estimation of the associated static
parameters based on data from an exponential family, i.e. the memory and
computing time of our method does not grow with time.
The inferential procedure derived in this paper is outlined below:

\begin{enumerate}
\item \label{item:2} For a given spatial range, $\phi$, we develop an online
  algorithm (Algorithm~\ref{alg:gibbs1}) for sampling from the filtering
  distribution and the posterior distribution of the other parameters.
  Our algorithm takes advantage of the skewness of the filtering
  distribution (see Lemma~\ref{prop:skewed}) and produces non-degenerate
  samples.
  
\item We develop an online estimate of the Bayes factors corresponding to a
  finite set of $\phi$ values which we then maximize to estimate $\phi$.
  The asymptotic validity of our method is established in
  Theorem~\ref{thm:est_phi_via_PF}.

\item Given an estimate of $\phi$, we derive a novel importance resampling
  algorithm (Algorithm~\ref{alg:pf_phi}) for estimating the latent process
  and the other parameters in an online fashion.
\end{enumerate}

The rest of the paper is organized as follows.
Section~\ref{sec:description} discusses the problem formulation and
displays preliminary results related to our model, including the derivation
of the skew-normal proposal and the proposed algorithm for sampling for
fixed $\phi$. Next, Section~\ref{subsec:phi} presents the main contribution
of this manuscript and considers the estimation of the spatial
correlation via our novel online implementation of the empirical Bayes
technique. In Section~\ref{sec:app} we illustrate the application of the
proposed method for online monitoring of radiation. Section~\ref{sec: disc}
offers a summary and discusses future research directions based on our
technique. The proofs to the results presented in the paper are provided in
the Appendix. In the Supplementary Materials for this article we present our
simulation study for assessing the performance of the proposed method,
including a comparison against a typical offline method.


\section{Problem formulation and preliminary results}
\label{sec:description}

\subsection{Model}
\label{sec:model}

We adopt a hierarchical, dynamical spatiotemporal model with the
observation process $Y$ having conditional independent components from an
exponential family distribution given a latent Gaussian spatiotemporal
field $X$.

Let $\mathbb{S}$ denote the continuous spatial domain of interest and let
$\mathbb{T} = \{0,1,2,\ldots\}$ be the temporal domain. The process model
consists of a spatiotemporal Gaussian process $X$ on
$\mathbb{S} \times \mathbb{T}$ having an autoregressive structure, that is,
for every finite collection of spatial locations in $\mathbb{S}$, say of
size $n$, the value of $X$ at the $n$ locations at time $t$, $\xbf_t$, is
described by a perturbation of its values at time $t-1$, $\xbf_{t-1}$, as
follows
\begin{equation}
  \label{eqn:dynamics}
  \xbf_0 = G_0 \beta + \sigma \epsilon_0,\quad
  \xbf_t = G_t \beta +
  \alpha (\xbf_{t-1}-G_{t-1} \beta) + \sigma \epsilon_{t},\ t=1,2,\ldots
\end{equation}
where the driving noise is a normally distributed isotropic spatial
process, $\epsilon_{t} \sim N_n (0, R(\phi))$, $G_t$ defines the
$n \times m$ matrix of covariates for each time $t$ associated with an
$m \times 1$ parameter vector $\beta$, $\sigma$ is the diffusion
coefficient and $\phi$ denotes the range parameter with spatial correlation
matrix $R = R(\phi)$. Our model assumes linear transition in time. In
general, $\alpha$ is an $n \times n$ matrix of redistribution weights,
however, as \cite{CrWi} argue in Section~7.2, this can be challenging to
estimate at early times. A simpler model that allows stationarity is to let
$\alpha$ be a scalar, where $|\alpha| < 1$. This implies homogeneous
transition between the components of the state process. The theory
developed in this paper applies to either matrix or scalar $\alpha$ but the issues
we wish to address can be illustrated more clearly when $\alpha$ is a
scalar so we will focus on this case for the remainder of this paper.

The observation process $Y$ is defined on $\mathbb{S} \times \mathbb{T}$
and is assumed to have conditionally independent components given the
latent spatiotemporal process $X$ with distribution from an exponential
family. In other words, let $y_{i,t}$ denote the value of the observation
process at location $i$ and time $t$, and let $x_{i,t}$ denote the value of
the corresponding spatiotemporal process. Then
\begin{equation}
  \label{eqn:obs}
  p(y_{i,t} | x_{i,t}) \propto
  \exp\{y_{i,t} g(x_{i,t}) - \tau_{i,t} b(x_{i,t}) \},
\end{equation}  
where under the usual regularity assumptions for exponential families the
mean of the distribution in~\eqref{eqn:obs} is $h({x}_{i,t})$ with
$h(\cdot)$ being the inverse link function, $g(\cdot)$ and $b(\cdot)$ are
known functions and $\tau_{i,t}$ is a known scalar associated with the
underlying distribution of the data. Further, given $x_{i,t}$, $y_{i,t}$ is
independent of every other component of $Y$.
The main advantage in using the general  state space representation
of a dynamic spatiotemporal process is that we do not need to rely on the
normality assumption for the observation process and thus nonlinear and/or
non-Gaussian models could be taken into account.

For the parameters $(\beta,\alpha)$ we assume a normal prior, although the
methodology described here is valid for truncated normal or improper
uniform priors as well, and the variance coefficient $\sigma^2$ is assumed to
be distributed according to an inverse-gamma conjugate prior, i.e.
\begin{equation}
  \label{eq:2}
  \begin{gathered}
    \beta|\sigma^2 \sim N_m(Q_0^{-1} b_0,\sigma^2 Q_0^{-1}), \\
    \alpha|\sigma^2 \sim N(s_0^{-1} a_0,\sigma^2 s_0^{-1}), \\
    \sigma^2 \sim IG \left( \frac{c_0}{2},\frac{r_0}{2} \right),
  \end{gathered}
\end{equation}
for suitable hyperparameters $b_0$, $Q_0$, $a_0$, $s_0$, $c_0$, and $r_0$.
To make the priors reasonably uninformative it is common to set
$Q_0 = q_0 I$, i.e. a diagonal with all diagonal elements equal to $q_0$,
and assign small values to $q_0$, $s_0$, $c_0$ and $r_0$. If $\alpha$ is a
matrix, as was discussed earlier, then matrix-normal prior is used instead.

\subsection{Bayesian inference}
\label{sec:bayesest}

We assume that observations from $n$ fixed spatial locations are available
at times $1,2,\ldots$, but we do not assume that all $n$ locations are
observed every time. We denote by
$\ybf_{1:t}= (\ybf_1,\ybf_2, \ldots,\ybf_t)$ the observations up to time
$t$ where each $\ybf_s$ is an $n$-dimensional vector, possibly containing
missing values. We denote by $\xbf_{0:t}= (\xbf_0,\xbf_1, \ldots,\xbf_t)$
the value of the spatiotemporal process at the same $n$ locations up to
time $t$, and $\theta=(\alpha,\beta,\sigma^2)$ denotes the temporal
parameters. The problem is to estimate the parameters $\theta$ and $\phi$,
as well as the latent process at time $t$, $\xbf_{t}$, from the available
data $\ybf_{1:t}$. The problem of estimating the spatiotemporal process at
locations different from the sampled locations becomes trivial after $\theta$,
$\phi$, and $\xbf_{0:t}$ are estimated.

For given $\phi$, we sample from the posterior distribution
$p(\xbf_{0:t},\theta| \ybf_{1:t}, \phi)$ by combining the particle filter
resampling method with a skewed proposal and the sufficient statistics
method of \cite{Svo} and \cite{Fea}. The setup of Section~\ref{sec:model}
allows sampling from the full conditional distribution of the parameters
$\theta$ via Gibbs sampling.

More precisely, the full conditional distribution of the
parameter $\beta$ is 
\begin{equation*}
  p(\beta|\xbf_{0:t}, \ybf_{1:t}, \alpha, \sigma^2, \phi) \propto p(\beta)
  \prod_{s=1}^t p(\xbf_s|\xbf_{s-1},\alpha, \beta,\sigma^2, \phi),
\end{equation*}
which is easily shown to be normal $
\beta | (\xbf_{0:t},\ybf_{1:t},\alpha,\sigma^2,\phi) \sim N_m(Q_t^{-1}
b_t,\sigma^2 Q_t^{-1})$, where
\begin{equation}
  \label{eq:3}
  \begin{aligned}
    Q_t &= Q_0 + G_0' R^{-1} G_0 + \sum_{s=1}^t (G_s- \alpha G_{s-1})' R^{-1} (G_s- \alpha
    G_{s-1}) , \\
    b_t &= b_0 + G_0' R^{-1} \xbf_0 + \sum_{s=1}^t (G_s - \alpha G_{s-1})'R^{-1} (\xbf_s- \alpha
    \xbf_{s-1}) ,
  \end{aligned}
\end{equation}
and similarly for $\alpha$ and $\sigma^2$.
Note that the full conditional distributions of
$\theta=(\alpha, \beta, \sigma^2)$ depend on some sufficient quantities,
$u_t=u_t(\xbf_{0:t},\phi)$, which are updated recursively. (We use the
term ``sufficient quantities'' instead of ``sufficient statistics'' because
they depend on the unknown parameter $\phi$.) For example, to
update $\beta$, from~\eqref{eq:3}, we need to keep a record of the sums
$\sum G_s' R^{-1} G_s$, $\sum G_s' R^{-1} G_{s-1}$,
$\sum G_s' R^{-1} \xbf_s$, $\sum G_s' R^{-1} \xbf_{s-1}$, and
$\sum G_{s-1}' R^{-1} \xbf_{s}$ where the summation is over
$s = 1, \ldots, t$. Having stored the sufficient quantities
$u_{t-1}(\xbf_{0:t-1},\phi)$ at time $t-1$, we update them by adding the
corresponding terms at time $t$, i.e.
$u_{t}(\xbf_{0:t},\phi) = \mathcal{U}(u_{t-1}(\xbf_{0:t-1},\phi),
\xbf_{t},\phi)$.

To sample $(\xbf_t,\theta)$ we take marginal samples from
$p(\xbf_{0:t},\theta|\ybf_{1:t},\phi)$. The key element in the process for
doing so, given a sample from $p(\xbf_{0:t-1}|\ybf_{1:t-1},\phi)$, is
\begin{equation*}
  p(\xbf_{0:t},\theta|\ybf_{1:t},\phi) \propto p(\ybf_t|\xbf_t)
  p(\theta|\xbf_{0:t-1},\phi) p(\xbf_t|\xbf_{t-1},\theta,\phi)
  p(\xbf_{0:t-1}|\ybf_{1:t-1},\phi) ,
\end{equation*}
which suggests a separate, two-step, update for $\theta$ and $\xbf_t$. In
the first step we sample $\theta$ from
$p(\theta|\xbf_{0:t-1},\phi) = p(\theta|u_{t-1}(\xbf_{0:t-1},\phi))$.
Because this distribution is not available in closed form but the full
conditional distributions for each component of $\theta$ given the other
components are, sampling is done by running a few Gibbs iterations for each
component of $\theta$. The final $\theta$ obtained at the end of the
sequence of Gibbs iterations, together with $\xbf_{t-1}$ and $\ybf_t$ are
used to sample $\xbf_t$. To that end, let
$q(\xbf_t|\xbf_{t-1},\ybf_t,\theta,\phi)$ be a proposal distribution which
generates $N$ particles $\tilde\xbf^{(1)}, \ldots, \tilde\xbf^{(N)}$. Each
particle carries a weight proportional to
\begin{equation}
  \label{weights}
  w^{(i)} =
  \frac{p(\ybf_t|\tilde\xbf_t^{(i)}) p(\tilde\xbf_t^{(i)}|\xbf_{t-1},\theta,\phi)}
  {q(\tilde\xbf_{t}^{(i)}|\xbf_{t-1},\ybf_{1:t},\theta,\phi)}.
\end{equation}
The sample $\xbf_t$ is chosen from among the
$\tilde\xbf^{(1)}, \ldots, \tilde\xbf^{(N)}$ with weights proportional to
$w^{(1)},\ldots,w^{(N)}$.
Algorithm~\ref{alg:gibbs1} outlines the steps from this procedure.

\begin{algorithm}[!htb]
  \caption{One step sampling  for fixed $\phi$ at time $t$.}
  \label{alg:gibbs1}
  \begin{algorithmic}[1]
    \item[\textbf{Input:}] $\ybf_t$; \\
    Sample $\xbf_{t-1} \sim p(\xbf_{t-1}|\ybf_{1:t-1},\phi)$;\\
    Sufficient quantities $u_{t-1} = u_{t-1}(\xbf_{0:t-1},\phi)$.
    
  \item[\textbf{Execute:}]

    \STATE Sample $\theta$ from $p(\theta|u_{t-1})$ by
    running a few Gibbs iterations.

    \STATE Compute the proposal
    $q(\xbf_t|\xbf_{t-1},\ybf_t,\theta,\phi)$.

    \STATE Sample
    $\tilde\xbf_t^{(1)}, \ldots, \tilde\xbf_t^{(N)} \sim
    q(\xbf_t|\xbf_{t-1},\ybf_t,\theta,\phi)$.

    \STATE Compute the weight $w^{(i)}$ according to~\eqref{weights} for
    $i=1,\ldots,N$.

    \STATE Sample index $j$ from $\{1,\ldots,N\}$ with weights proportional
    to $w^{(1)},\ldots,w^{(N)}$.

    \STATE Set $\xbf_t = \tilde\xbf_t^{(j)}$
    
    \STATE Update the sufficient quantities $u_t =
    \mathcal{U}(u_{t-1},\xbf_t,\phi)$. 

    \item[\textbf{Return:}] $\xbf_t$, $u_t$, $\theta$.
  \end{algorithmic}
\end{algorithm}

On the other hand, implementing the same approach for the estimation of the
range parameter $\phi$ is far from trivial and it cannot be updated using
sufficient quantities as before, making it impractical for online
applications. Furthermore, each update of $\phi$ requires the inversion of
a large matrix which could be computationally expensive when the dimension
of the random field, $n$, is large. Therefore a technique other than Gibbs
sampling (or in general an MCMC framework) is required. We circumvent this
problem by using a novel particle filter with an
online implementation of an empirical Bayes method. This technique is
treated next in Section~\ref{subsec:phi}.

\subsection{A skewed-normal proposal density}

A measure of the quality of the proposal distribution is
the effective sample size (ESS), defined as
\begin{equation*}
  \mathrm{ESS} = \frac{(\sum_i w^i_t)^2}{\sum_i (w^i_t)^2} .
\end{equation*}
It can take values between 1 and $N$ and is used for assessing the loss of
variance in the importance weights \citep[][Section~4.4]{RobertCasella}.
A value close to $N$ would mean that the samples are nearly equally
weighted and there is diversity among the samples so no samples are lost. A
value close to 1 would indicate that all but one sample will have weight
close to 0 and lead to the well-known problem of sample degeneracy.

Importance sampling methods allow flexibility in the choice of the proposal
distribution $q$ but as \cite{DGA}
point out the optimal proposal distribution in the sense that it minimizes the
variance of the importance weights is
\begin{equation}
  \label{eq:9}
  q(\xbf_{t}|\xbf_{t-1},\ybf_{1:t},\theta,\phi) =
  p(\xbf_{t}|\xbf_{t-1},\ybf_{t},\theta,\phi).
\end{equation}
Although not helpful by itself, equation~\eqref{eq:9} is still useful since
it can serve as a basis for deriving suboptimal proposal distributions. One
way of doing this approximates~\eqref{eq:9} by a multivariate Gaussian
distribution as was done by \cite{DGA} in the \emph{univariate} case. When
the state process is multivariate it is imperative to use a good proposal
density as this increases the effective sample size. In view of this, we
introduce next a novel importance density. First, Lemma~\ref{prop:skewed}
shows that the optimal proposal distribution is skewed when the observation
process has a skewed distribution. This suggests using a proposal
distribution which is skewed and motivates our use of the skewed-normal
distribution for this purpose.

\begin{lemma}
  \label{prop:skewed}
  Consider the stochastic dynamics in equation~\eqref{eqn:dynamics}
  and the observation process given in equation~\eqref{eqn:obs} which
  corresponds to data from a general exponential family. Then the optimal
  proposal distribution given in equation~\eqref{eq:9} is skewed when the
  likelihood of $y_{i,t}|x_{i,t}$ is skewed.
\end{lemma}
\begin{proof}
  See Appendix.
\end{proof}

Consider first a Gaussian approximation to~\eqref{eq:9}. Note that the
optimal proposal is
$p(\xbf_{t}|\xbf_{t-1},\ybf_{t},\theta,\phi) \propto p(\ybf_t|\xbf_t)
p(\xbf_t|\xbf_{t-1},\theta,\phi)$ and let
\begin{equation}
  \label{eq:1}
  f(\xbf_t) = -\log \{p(\ybf_t|\xbf_t) p(\xbf_t|\xbf_{t-1},\theta,\phi)\}.
\end{equation}
Next define
\begin{equation*}
  \hat \xbf_t = \argmin_{\xbf_t} f (\xbf_t) , \qquad
  \hat H_t = \nabla \nabla' f(\hat \xbf_t) . 
\end{equation*}
The Gaussian proposal is constructed by setting the mean equal to
$\hat{\xbf}_t$ and the variance to $\hat{H}_t^{-1}$.

To capture the skewness of the distribution we use a skewed-normal copula
correction to the Gaussian proposal. The probability density function (pdf)
of the univariate skewed-normal distribution is \citep{AC:SN}
\begin{equation*}
  \frac{2}{\omega}\, \psi\! \left( \frac{z-\xi}{\omega} \right) \Psi\! \left( a
    \frac{z-\xi}{\omega} \right)
\end{equation*}
where $\psi(\cdot)$ and $\Psi(\cdot)$ denote the pdf and cumulative
distribution function (cdf) of the standard normal distribution
respectively. The parameters $\xi$, $\omega >0$, and $a$ correspond to the
location, scale, and skewness parameter respectively.

To derive the skewed-normal corrections we expand the marginals
of~\eqref{eq:1} to third order terms and match the first three moments to
the skewed-normal distribution. For details see Appendix~B of
\cite{rue2009approximate}. Let $\tilde\xbf_t$ be a sample from the Gaussian
approximation to~\eqref{eq:9}. The idea is to transform $\tilde\xbf_t$
marginally using the skewed-normal correction. \cite{ferkingstad2015}
propose two copula corrections which we also use here: a mean-only skewness
correction where the proposal distribution remains Gaussian but the mean is
corrected using the skewed-normal approximations to the marginals; and a
mean-plus-skewness correction where the particles are sampled from the
Gaussian approximation and then are marginally transformed using the
skewed-normal approximation. 
In our simulation study (see Supplementary Materials, Section~1) we find
that the two skewed-normal proposals have similar ESS but the Gaussian
proposal has significantly lower ESS. We therefore recommend the mean-only
corrected proposal because it is simpler than the mean-plus-skewness
correction, while it also has better ESS than the Gaussian proposal.

\section{A methodology for online estimation and prediction}
\label{subsec:phi}

In this section we present an empirical Bayes approach for the estimation
of the range parameter $\phi$. Unlike the parameter
$\theta=(\alpha,\beta,\sigma^2)$, it is not possible to include $\phi$ as
an extra step in the Gibbs algorithm without sacrificing the online feature
of the method since the sufficient quantities for the update of $\theta$
depend on $\phi$ and consequently they must be recomputed from time one at
every update of $\phi$. Instead we adopt a novel empirical Bayes method in
order to estimate $\phi$ in a way that is similar in spirit to \cite{Doss}.
Another argument in favor of the empirical Bayes approach instead of a full
Bayesian approach is that it is unclear how a suitable prior for $\phi$
should be chosen. \cite{BOS} discuss some objective priors in the case of
Gaussian responses. However for non-Gaussian data these priors, and indeed
any improper prior, result in an improper posterior for $\phi$ \citep{CMW}.
When it comes to online inference, it is unclear how a fully Bayesian
approach would be implemented. If a Monte-Carlo algorithm is used, the
sufficient quantities will be computed for those $\phi$ values in the
Monte-Carlo sample only. This restricts the $\phi$ values at subsequent
times to only those which were sampled at all previous time points,
something that is undesirable. The goodness-of-fit of the empirical Bayes
method for estimating the range parameter has been demonstrated in the case
of the spatial-only model by \cite{roy16}. Extending it to an online
version requires careful treatment of the sufficient quantities needed to
compute the Bayes factors. Theorem~\ref{thm:est_phi_via_PF} presents the
main result of this section. The approach discussed below may be also
viewed as equivalent to the maximum likelihood estimation for $\phi$ after
integrating out the parameter $\theta$.

We consider first the marginal density $p(\ybf_{1:t}|\phi)$ and define
the estimator for $\phi$ at time $t$ given data $\ybf_{1:t}$ by
\begin{equation}
  \label{eq:14}
  \hat \phi_t = \argmax_\phi p(\ybf_{1:t}|\phi),
\end{equation}
where
\begin{equation}
  \label{eq:15}
  p(\ybf_{1:t}|\phi) = \int p(\ybf_{1:t},\xbf_{0:t},\theta|\phi)
  d (\xbf_{0:t},\theta) . 
\end{equation}
In general, the integral in~\eqref{eq:15} has no closed form  and
thus a numerical approximation must be employed. Define the sequential Bayes factor
between $\phi$ and $\tilde \phi$ with respect to the data $\ybf_{1:t}$ by
\begin{equation*}
  B_{1:t}(\phi;\tilde{\phi}) =
  \frac{p(\ybf_{1:t}|\phi)}{p(\ybf_{1:t}|\tilde{\phi})}.
\end{equation*}
Note the dependence of the Bayes factor on the whole data sequence
$\ybf_{1:t}$. Then, for a fixed parameter $\tilde \phi$, \eqref{eq:14} is
equivalent to
\begin{equation*}
  \hat \phi_t = \argmax_\phi  B_{1:t}(\phi;\tilde{\phi}).
\end{equation*}
Furthermore, the sequential Bayes factor,
$B_{1:t}(\phi;\tilde{\phi})$ in a filtering framework is computed as
follows:
\begin{equation}
  \label{eq:16}
  \begin{aligned}
    B_{1:t}(\phi;\tilde{\phi}) &= \int
    \frac{p(\ybf_{1:t},\xbf_{0:t},\theta|\phi)}
    {p(\ybf_{1:t},\xbf_{0:t},\theta|\tilde{\phi})}
    p(\xbf_{0:t},\theta|\ybf_{1:t},\tilde{\phi}) d(\xbf_{0:t},\theta) \\
    &= \int
    \frac{p(\ybf_{1:t}|\xbf_{0:t}) p(\xbf_{0:t}|\theta,\phi) p(\theta)}
    {p(\ybf_{1:t}|\xbf_{0:t}) p(\xbf_{0:t}|\theta,\tilde{\phi}) p(\theta)}
    p(\xbf_{0:t},\theta|\ybf_{1:t},\tilde{\phi}) d(\xbf_{0:t},\theta) \\
    &= \int
    \frac{p(\xbf_{0:t}|\theta,\phi)}{p(\xbf_{0:t}|\theta,\tilde{\phi})}
    p(\xbf_{0:t},\theta|\ybf_{1:t},\tilde{\phi}) d(\xbf_{0:t},\theta) .
  \end{aligned}
\end{equation}
A naive approach for estimating $\phi$ relying on equation \eqref{eq:16}
would be to obtain a large sample for $(\xbf_{0:t},\theta)$ from
$p(\xbf_{0:t},\theta|\ybf_{1:t},\tilde{\phi})$ using
Algorithm~\ref{alg:gibbs1}, and to approximate~\eqref{eq:16} by Monte-Carlo
integration; call the result $\hat B_{1:t}(\phi;\tilde\phi)$. Then an estimate
would be obtained by maximizing $\hat B_{1:t}(\phi;\tilde\phi)$ over
$\phi$.

\begin{remark} \label{naMC} There are several issues concerning the above
  naive approach that need to be addressed:
\begin{enumerate}
\item \label{item:3} Unless $\hat \phi_t$ and $\tilde \phi$ are
  close to one another, the Monte-Carlo approximation will have a large error
  and in this case the estimate may not be accurate no matter how large the
  Monte-Carlo sample is.
\item \label{item:4} In order to compute $p(\xbf_{0:t}|\theta,{\phi})$ for
  any $\phi$ at time-point $t$, we need $p(\xbf_{0:t-1}|\theta,{\phi})$ for
  the same $\phi$ at time $t-1$ something that we could not anticipate
  prior to time $t$.
\item \label{item:5} After obtaining $\hat \phi_t$ we need to run
  Algorithm~\ref{alg:gibbs1} once more in order to update $\xbf_t$ and
  $\theta$ conditional on $\hat\phi_t$. However the algorithm requires
  samples from $\xbf_{0:t-1}|\ybf_{1:t-1},\hat\phi_t$  which
  is unknown at time $t-1$.
\end{enumerate}
\end{remark}

Bypassing the issues raised in Remark~\ref{naMC}, our strategy for
resolving issue~\ref{item:3} is to replace the importance density
in~\eqref{eq:16} by a mixture over a set of $\phi$ values instead of a
single fixed $\tilde \phi$. This is demonstrated in our simulation study
(see Supplementary Materials, Section~2) where we show that the naive
approach can give biased estimates for a badly chosen $\tilde{\phi}$.
For~\ref{item:4} we only evaluate the Bayes factors over a dense grid
$\Phi$ which covers reasonable values of $\phi$ and keeps a record of the
sufficient quantities needed to evaluate $p(\xbf_{0:t}|\theta,{\phi})$ at
the next time. To resolve~\ref{item:5} we resample the available samples
from the mixture with appropriate weights. In the remainder of this section
we expand on these ideas. Algorithm~\ref{alg:pf_phi} puts them together.

Consider a set $\Phi_K = \{\phi_1,\ldots,\phi_K\}$ such that
$\tilde\phi \in \Phi_K$, is sufficiently spread-out over a range of
interesting values of $\phi$. The meaning of ``interesting values of
$\phi$'' is well defined in our context: the range parameter is a scaling
factor of the spatial distances within the domain of interest which define
a possible range for $\phi$. Suppose $(\xbf_{0:t}^{(l,k)},\theta^{(l,k)})$,
$k=1,\ldots,K$, $l=1,\ldots,L_k$ are samples from
$p(\xbf_{0:t}, \theta | \ybf_{1:t}, \phi_k)$. The augmented sample can be
seen as drawn from the mixture distribution 
\begin{equation}
  \label{eq:pmix}
  p_{\text{mix}}(\xbf_{0:t},\theta|\ybf_{1:t},\Phi_K,\Lambda_K) =
  \sum_{k=1}^K \lambda_k p(\xbf_{0:t},\theta|\ybf_{1:t},\phi_k) ,
\end{equation}
where $\lambda_k = L_k/(\sum L_{k'})$ and
$\Lambda_K = \{\lambda_1,\ldots,\lambda_K\}$. Let
$b_t^k=B_{1:t}(\phi_k;\tilde \phi)$. Then $b_t^k$ can be estimated by
maximizing the so-called reverse logistic log-likelihood \citep{Gey}
\begin{equation*}
  \ell(\mathbf{b}_t) = \sum_{k=1}^K \sum_{l=1}^{L_k} \log
  \frac{\lambda_k
    p(\xbf_{0:t}^{(k,l)},\theta^{(k,l)}|\ybf_{1:t},\phi_k)}
  {p_{\text{{mix}}}(\xbf_{0:t}^{(k,l)},\theta^{(k,l)}|\ybf_{1:t},  
    \Phi_K,\Lambda_K)} .
\end{equation*}
Furthermore, let $\hat{b}_t^k$ denote the estimate for $b_t^k$. Then, the following sum
\begin{equation}
  \label{eq:37}
  \hat{B}_{1:t}(\phi;\tilde \phi) =  \sum_{k=1}^K \sum_{l=1}^{L_k}
  \frac{p(\xbf_{0:t}^{(k,l)}|
    \theta^{(k,l)},\phi)} 
  {\sum_{k'} {L_{k'}}/{\hat b^{k'}_t}  
    p(\xbf_{0:t}^{(k,l)}|\theta^{(k,l)},\phi_{k'})},
\end{equation}
estimates $B_{1:t}(\phi;\tilde{\phi})$. The key Theorem \ref{thm:est_phi_via_PF} summarizes this property.

\begin{theorem}
  \label{thm:est_phi_via_PF}
  Consider a coarse grid $\Phi_K = \{\phi_1, \phi_2, \ldots, \phi_K \}$,
  where the grid points, $\phi_k$, $k=1,\ldots, K$, are spaced across the
  parameter space for $\phi$. Suppose that for $k=1,\ldots,K$, we draw
  samples $(\xbf^{(k,l)}_{1:t},\theta^{(k,l)})$,
  $l=1,\ldots,L_k$ from the distribution
  $p(\xbf_{0:t},\theta| \phi_k, \ybf_{1:t})$ for $\phi_k \in \Phi_K$. Then,
  for an arbitrarily fixed pair $(\phi, \tilde{\phi})$, the estimate,
  \begin{equation}
    \hat{B}_{1:t}(\phi;\tilde \phi) \stackrel{\mbox{a.s.}}{\longrightarrow}
    B_{1:t}(\phi; \tilde{\phi}), \; \; L_k \to \infty,
  \end{equation}
  where $ \hat{B}_{1:t}(\phi;\tilde \phi)$ is given by equation \eqref{eq:37}.
\end{theorem}
\begin{proof}
  See Appendix.
\end{proof}

The likelihood in the numerator and denominator of~\eqref{eq:37} must be
computed for the whole history of samples $\xbf_{0:t}^{(k,l)}$ for the new
$\theta^{(k,l)}$ and for different values of $\phi$. This is not as straightforward as a product of the prior
$p(\xbf_{0:t-1}|\theta,\phi)$ times the transition
$p(\xbf_{t}|\xbf_{t-1},\theta,\phi)$ since in online implementations we
cannot anticipate the value of $\theta$ and $\phi$ at time $t-1$. If $\phi$
is held fixed, then $p(\xbf_{0:t}|\theta,\phi)$ can be written in terms of
sufficient quantities as before which do not depend on $\theta$ but do
depend on $\phi$.

To this end, let $z_{t-1}^{(\phi,k,l)}$ denote the aforementioned
sufficient quantities at time $t-1$ for the $(k,l)$th sample. These
are updated at time $t$ by
\begin{equation*}
  z_{t}^{(\phi,k,l)} =
  \mathcal{Z} \left(z_{t-1}^{(\phi,k,l)},\xbf_{t}^{(k,l)},\phi\right),
\end{equation*}
and the joint likelihood is expressed as a function of $z_{t}^{(\phi,k,l)}$
and $\theta^{(k,l)}$, i.e.
\begin{equation*}
  p(\xbf_{0:t}^{(k,l)}|\theta^{(k,l)},\phi) = \mathcal{P}
  \left(z_{t}^{(\phi,k,l)}, \theta^{(k,l)} \right) .
\end{equation*}
  
In practice, we do not consider the entire parameter space for $\phi$ but a
fine discretization of it. More precisely, we augment the coarse grid, $\Phi_K,$
with the finer grid, say $\Phi$, i.e. $\Phi_K \subset \Phi$, and compute
the sufficient quantities $z_{t}^{(\phi,k,l)}$, and consequently
$\hat B_{1:t}(\phi;\tilde\phi)$, only for those $\phi \in \Phi$. Then, we
estimate $\phi$ by
\begin{equation*}
  \hat{\phi}_t = \argmax_{\phi \in \Phi}
  \hat B(\phi;\tilde \phi) .
\end{equation*}
The drawback is the loss of precision in  estimation but the benefit is
that the required computing memory remains fixed. From experience, we find that a
small bias in the value of $\phi$ does not affect prediction or the
estimation of the other parameters. In particular in the simulation study
presented in Section~2 of the Supplementary Materials we
find that the parameters $\alpha$ and $\beta$ are immune to the possible
bias in $\hat{\phi}_t$.

The empirical Bayes approach proceeds with the update of the estimate for
$(\xbf_t,\theta)$ using the estimate $\hat\phi_t$.
These estimates are obtained as a weighted sum of the existing samples
$(\xbf_{0:t}^{(k,l)},\theta^{(k,l)})$, $k=1,\ldots,K$, $l=1,\ldots,L_k$.
{The distribution of the existing samples is the mixture
  distribution~\eqref{eq:pmix}. These samples can be scaled with reference
  to the distribution conditioned on $\phi = \hat{\phi}_t$ using the
  following importance weights}
\begin{align} \label{weights_after_EB}
  v_t^{(k,l)} &= \frac{p(\xbf_{0:t}^{(k,l)},\theta^{(k,l)}|\ybf_{1:t},\hat
    \phi_t)} {p_{\text{mix}}(\xbf_{0:t}^{(k,l)},\theta^{(k,l)}
    |\ybf_{1:t},\Phi_K,\Lambda_K)} 
  = \frac{p(\xbf_{0:t}^{(k,l)},\theta^{(k,l)}|\ybf_{1:t},\hat \phi_t)}
  {\sum_{k'} \lambda_{k'}
    p(\xbf_{0:t}^{(k,l)},\theta^{(k,l)}|\ybf_{1:t},\phi_{k'})} \notag \\
  &= \frac{p(\ybf_{1:t}|\xbf_{0:t}^{(k,l)}) p(\xbf_{0:t}^{(k,l)}|\theta^{(k,l)},
    \hat \phi_t) p(\theta^{(k,l)}) / p(\ybf_{1:t}|\hat\phi_t)} {\sum_{k'}
    \lambda_{k'} p(\ybf_{1:t}|\xbf_{0:t}^{(k,l)})
    p(\xbf_{0:t}^{(k,l)}|\theta^{(k,l)},\phi_{k'}) p(\theta^{(k,l)}) /
    p(\ybf_{1:t}|\phi_{k'})} \notag \\
 & = \frac{p(\xbf_{0:t}^{(k,l)}|\theta^{(k,l)},
    \hat \phi_t) / B_{1:t}(\hat\phi_t ; \tilde \phi)} {\sum_{k'}
    \lambda_{k'} 
    p(\xbf_{0:t}^{(k,l)}|\theta^{(k,l)},\phi_{k'}) /
    B_{1:t}(\phi_{k'};\tilde \phi)} . 
\end{align}
In practice $B_{1:t}(\hat\phi_t ; \tilde \phi)$ and $B_{1:t}(\phi_{k'} ;
\tilde \phi)$ in equation \eqref{weights_after_EB} are replaced by their
estimates which are already available. Then, we obtain the
  estimates for the latent state process and the remaining parameters by
\begin{equation*}
\hat \xbf_t = \sum_{k=1}^K \sum_{l=1}^{L_k} \bar v_t^{(k,l)}
    \xbf_t^{(k,l)}, \quad \hat
    \theta_t = \sum_{k=1}^K \sum_{l=1}^{L_k} \bar v_t^{(k,l)} \theta^{(k,l)},
\end{equation*}
where $\bar v_t^{(k,l)}$ is the normalized version
of~\eqref{weights_after_EB}. This is the final step at time $t$. The main
algorithm of this paper which shows how to combine
Algorithm~\ref{alg:gibbs1} with the online empirical Bayes for the
estimation of $\phi$ is displayed in Algorithm~\ref{alg:pf_phi}.

\begin{algorithm}[!tb] 
  \caption{Main estimation and prediction algorithm at time $t$. }
  \label{alg:pf_phi}
  \begin{algorithmic}[1]
    \item[\textbf{Input:}] $\ybf_{t}$; \\
    Samples $\xbf_{t-1}^{(k,1:L_k)} \sim
    p(\xbf_{t-1}|\ybf_{1:t-1},\phi_k)$, $k=1,\ldots,K$; \\
    Sufficient quantities $u_{t-1}^{(k,1:L_k)}$, $k=1,\ldots,K$;\\
    Sufficient quantities $z_{t-1}^{(\phi,k,1:L_k)}$, $\phi \in \Phi$, $k =
    1,\ldots,K$. \\ 
    
    \item[\textbf{Execute:}]
    
    \FORCONC {$k \in \{1,\ldots, K\}$}
    
    \FORCONC {$l = 1,\ldots,L_k$}

    \STATE Sample $l'$ uniformly in $\{1,\ldots,L_k\}$.

    \STATE Call Algorithm~\ref{alg:gibbs1} with input $\ybf_t$,
    $\xbf_{t-1}^{(k,l')}$, $u_{t-1}^{(k,l')}$,
    and output $\theta^{(k,l)}$, $\xbf_t^{(k,l)}$, $u_t^{(k,l)}$.

    \FORCONC {$\phi \in \Phi$}
    
    \STATE Update the sufficient quantities $z_{t}^{(\phi,k,l)} =
    \mathcal{Z} \left(z_{t-1}^{(\phi,k,l)},
      \xbf_{t}^{(k,l)},\phi \right)$.
    
    \STATE Compute $p(\xbf_{0:t}^{(k,l)}|\theta^{(k,l)},\phi) =
    \mathcal{P} \left(z_{t}^{(\phi,k,l)}, \theta^{(k,l)} \right)$.
    
    \ENDFOR \ $\phi$
    
    \ENDFOR \ $l$

    \ENDFOR \ $k$

    \STATE Call the reverse logistic regression algorithm with input
    $\{p(\xbf_{0:t}^{(k,1:L_k)}|\theta^{(k,1:L_k)},\phi): \phi \in
    \Phi, k=1,\ldots,K \}$ and output $\hat{\mathbf{b}}_t$.

    \STATE Compute $\hat B_{1:t}(\phi;\tilde \phi)$, $\phi \in \Phi$
    using~\eqref{eq:37}.

    \STATE Set $ \hat\phi_t = \argmax_{\phi \in \Phi} \hat
    B_{1:t}(\phi;\tilde \phi)$.

    \STATE Compute importance weights $v_t^{(k,l)}$ according to equation
    \eqref{weights_after_EB} and normalize them to get $\bar v_t^{(k,l)}$.

    \STATE Set $\hat \xbf_t = \sum_{k=1}^K \sum_{l=1}^{L_k} \bar v_t^{(k,l)}
    \xbf_t^{(k,l)}$, $\hat
    \theta_t = \sum_{k=1}^K \sum_{l=1}^{L_k} \bar v_t^{(k,l)} \theta^{(k,l)}$.
    
  \item[\textbf{Return:}] $\hat \xbf_t$, $\hat \theta_t$, $\hat \phi_t$, \\
    $\xbf_{t}^{(k,1:L_k)}$, $\theta^{(k,1:L_k)}$, $u_{t}^{(k,1:L_k)}$,
    $k=1,\ldots,K$,\\ 
    $z_{t}^{(\phi,k,1:L_k)}$, $\phi \in \Phi$, $k = 1,\ldots,K$.
  \end{algorithmic}
\end{algorithm}

We evaluated the performance of our method in a simulation study which we
include in the Supplementary Materials with this article. Our study shows
that the obtained parameter estimates are unbiased and consistent. As more
data are assimilated, the credible intervals obtained from the Monte-Carlo
samples become narrower as expected with rate in the order of the square
root of the number of the elapsed time. The posterior distributions
obtained by our method, conditioned on the data $\ybf_{1:t}$, are compared
against those derived from a typical offline, fully Bayesian, MCMC method.
Indeed, the two distributions match which indicates the Monte-Carlo samples
are taken from the correct distribution. Overall, this study verifies our
theoretical conclusions and justifies the use of our method for online
inference.

\section{Example: Spatiotemporal monitoring of the Cs-137 isotope}
\label{sec:app}

The Fukushima Daiichi nuclear disaster was a catastrophic failure at the
Fukushima Nuclear Power Plant on 11 March 2011, resulting in a meltdown of
three of the plant's six nuclear reactors. The failure occurred when the
plant was hit by the tsunami following an earthquake. The Japanese
authorities started to collect data about the radioactive material released
from the power station which were reported to the International Atomic
Energy Agency (IAEA). At an early stage of the accident, online methods
were needed to incorporate new measurements in real-time. The data analyzed
in this paper consist of daily measurements of radioactive decay for the
Caesium-137 (Cs-137) isotope found on leaves collected between 16 March
2011 and 26 December 2011. The measurements were collected from different locations and the number of nuclear decays of the
isotope in one second were counted. We refer the reader to the IAEA
relevant website \verb|https://iec.iaea.org/fmd| for more information and
access to the datasets. 

\begin{figure}[!tb]
  \centering
  \includegraphics[width=.7\linewidth]{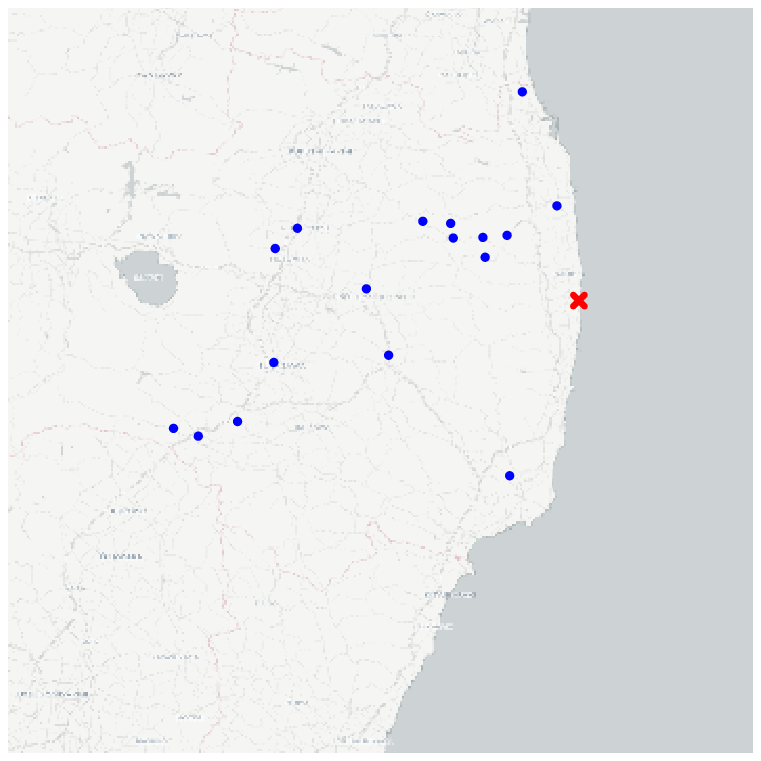}
  \caption{Sampled locations of the Cs-137 isotope, shown by a
    $\bullet$. The location of the power plant is shown by a $\times$.}
  \label{fig:4}
\end{figure}

The samples were taken at $n=17$ distinct locations across $T=146$ days.
However some locations were sampled more than once on the same day so the
total of all the measurements were used for that day. An intercept term, a time trend,
and the distance from the power plant were used as covariates. Our aim is to
estimate the parameters $(\alpha,\beta,\sigma^2,\phi)$ as well as
predict the spatiotemporal field $\xbf_t$ at time $t$ from observations
$\ybf_{1:t}$. In other words the hidden spatiotemporal process is given by
\begin{align}
  \label{pro_fuk}
  \textbf{x}_t &= \beta_0 + \beta_1 g + \beta_2
  t + \eta_t, \notag \\ \eta_t &= \alpha \eta_{t-1} + \epsilon_t,
\end{align}
where $g$ is the distance from the station, $\epsilon_t \sim \mathcal{N}(0,
\sigma^2 R(\phi)),$ and $\mathrm{Corr}(x_{i,t},x_{j,t})= e^{-d_{ij}/\phi}.$
Moreover, the $i$th collected observation is conditionally distributed
according to, $$ y_{i,t}|x_{i,t} \sim \mbox{Poisson}(\tau_{i,t}
e^{x_{i,t}}),$$ where $\tau_{i,t}$ corresponds to the number of times that
location $i$ was sampled at day $t$. These locations are shown in
Figure~\ref{fig:4}.

The priors for $(\alpha,\beta,\sigma^2)$ were used as in
Section~\ref{sec:description} with the following parameters: $a_0 = 0$,
$s_0 = 0.1$, $b_0 = 0$, $q_0 = 0.01$, $d_0 = 0.1$, $e_0 = 0.1$. The fine
grid $\Phi$ for estimating $\phi$ consists of 51 equally spaced points in
$[0,0.1]$ and the coarse grid $\Phi_K$ consists of 7 equally spaced points
in [0.002,0.098]. Algorithm~\ref{alg:pf_phi} was used
for estimation and prediction with particle size $N=1000$, MCMC size
$L=500$ and Gibbs burn-in $B=100$.

\begin{figure}[!htb]
  \centering
  \includegraphics[width=\linewidth]{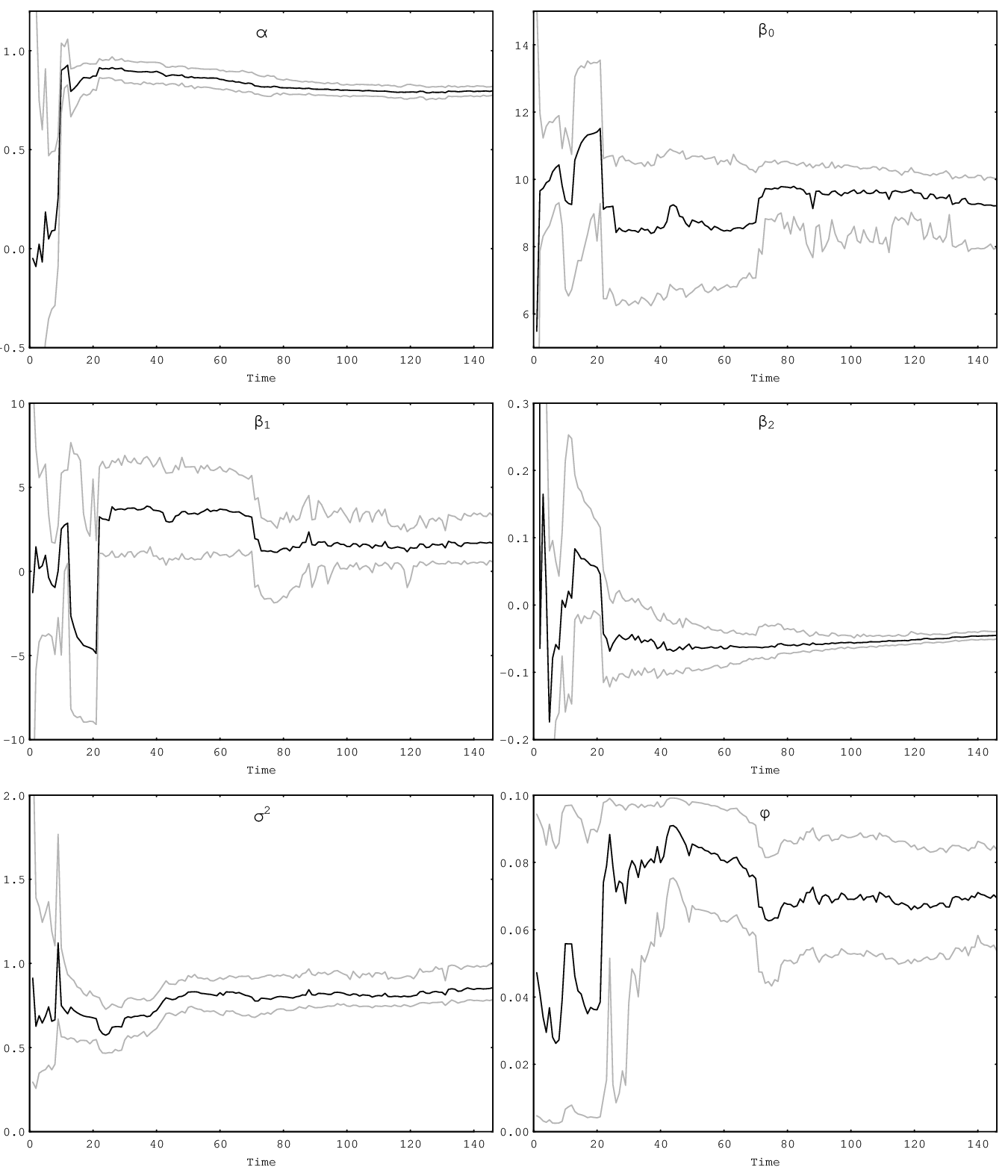}
  \caption{Estimates and 90\% credible intervals for the parameters
    of the Fukushima power
    plant example.}
  \label{fig:5}
\end{figure}

Figure~\ref{fig:5} shows the evolution of the parameter estimates in time.
There is an apparent ``jump'' in the parameter estimates for $\beta$ and
$\phi$ at around time $t=70$ after which the estimates become stable. A
closer examination of the data reveals that this may be due to a lower
trend after time 50 and to reduced sampling after time 40.

\begin{figure}
  \centering
  \includegraphics[width=\linewidth]{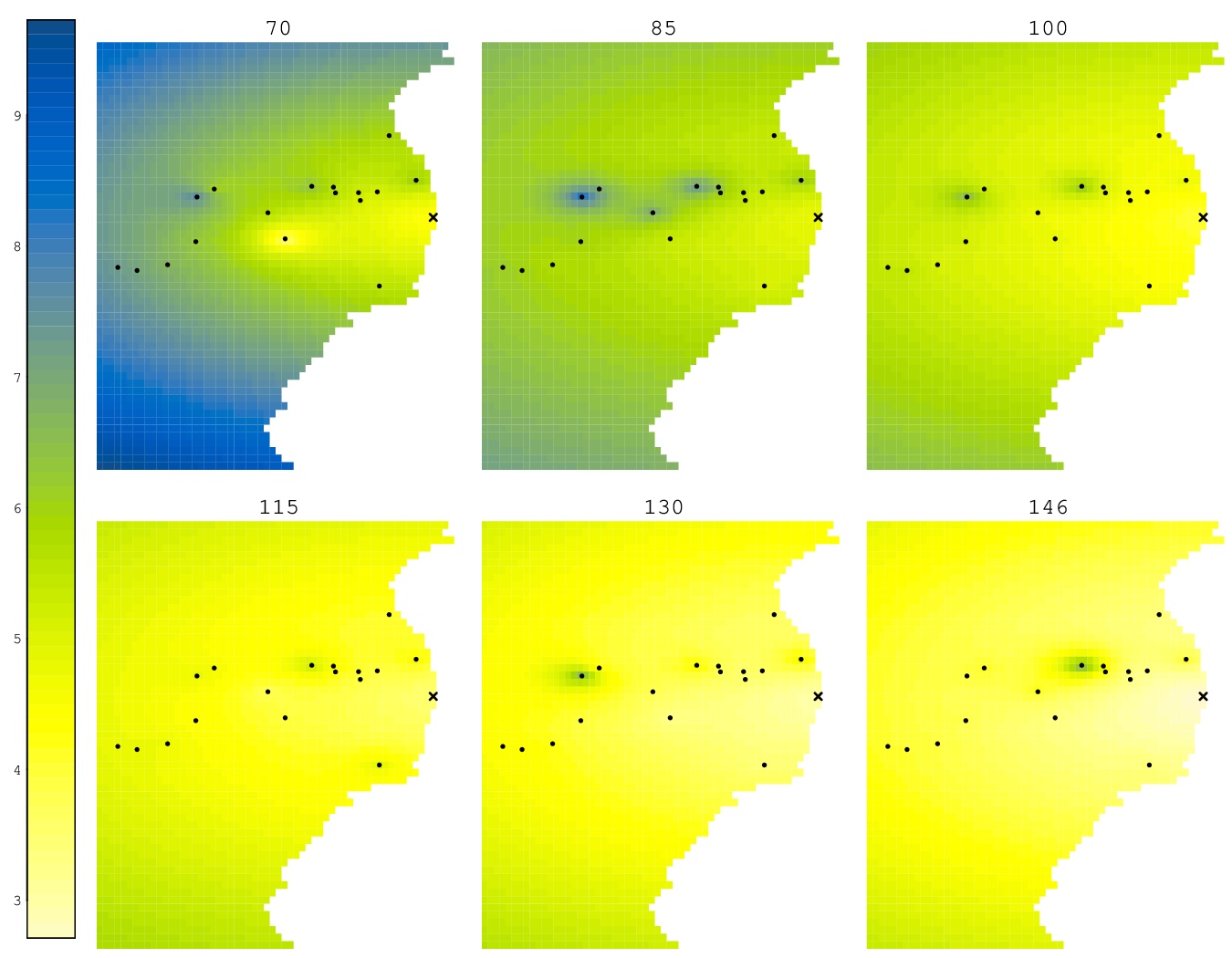}
  \caption{Predictions of the state process at different times for the
    Fukushima power plant example.}
  \label{fig:fuk_pred}
\end{figure}

\begin{figure}
  \centering
  \includegraphics[width=\linewidth]{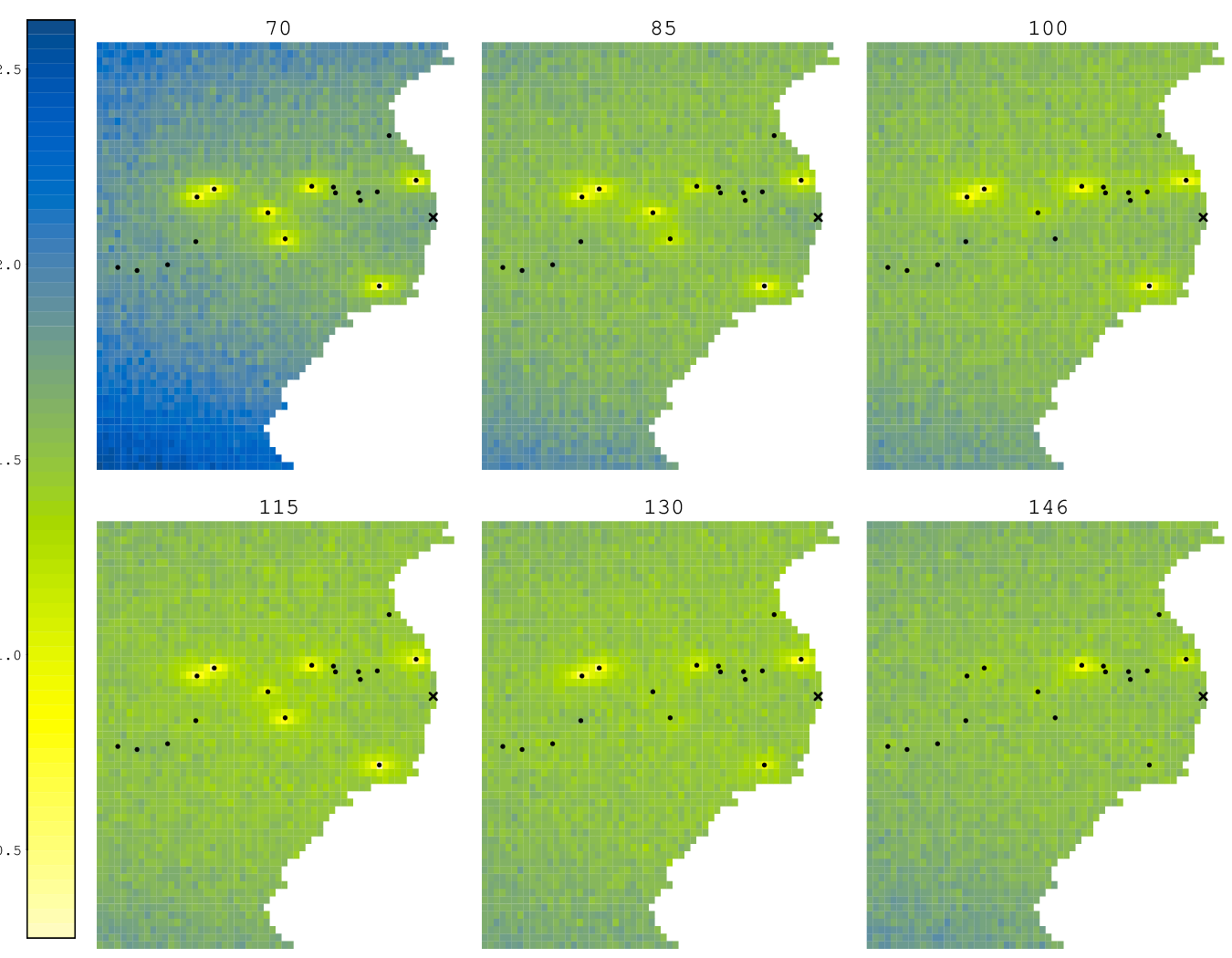}
  \caption{Prediction standard deviation of the state process at different
    times for the Fukushima power plant example.}
  \label{fig:fuk_predsd}
\end{figure}

Figure~\ref{fig:fuk_pred} shows the prediction at 2659 locations around the
sampling area for selected times. To sample from the unmonitored locations
we simulate from its conditional distribution given the samples at the
monitored locations and the parameters,
$p(\xbf_{t}^*|\xbf_{0:t},\theta, \phi)$, where $\xbf_{t}^*$ is the value of
the state process at the prediction locations. Figure~\ref{fig:fuk_predsd}
shows the standard deviation of our predictions. From the plots we can
identify some radiation hot-spots and an apparent decrease of radiation
over time.

\clearpage

\section{Summary and discussion}
\label{sec: disc}

In this paper we propose a method for online estimation and prediction of
dynamic spatiotemporal processes. We consider a latent Gaussian
autoregressive spatial process with data arriving sequentially in time with
distribution from an exponential family conditional on the spatiotemporal
process. Our model is expressed in terms of unknown parameters which are
estimated along with the latent process within an empirical Bayes
framework. We distinguish two types of parameters, the temporal parameters,
which have a full conditional distribution that can be written in terms of
sufficient quantities, and the spatial correlation parameters, which don't.
The spatial range parameter belongs to the latter type.

An algorithm is proposed for sampling from the filtering distribution of
the spatiotemporal process and the posterior distribution of those
parameters whose full conditional can be written in terms of sufficient
quantities. These sufficient quantities are updated when new samples are
taken for a fixed value of the range parameter, and because the filtering
distribution can be skewed, we show how to use the skew-normal distribution
to generate good candidate samples. The advantage of using sufficient
quantities is that the storage requirements do not increase in time.
These sufficient quantities depend on the spatial range parameter, thus,
they are computed at a fixed set of values of that parameter across
different times. Estimation of the range parameter is performed by
maximizing the Bayes factors over this fixed set. The Bayes factors are
estimated sequentially by importance sampling using the Monte Carlo
samples.

Our method was compared against a typical offline MCMC method which samples
from the posterior distribution of all parameters and the spatiotemporal
field. We find that the distribution of the samples from our method matches
the one obtained when the offline method is used. Finally, we demonstrate
the application of our method on radiation measurements from a nuclear
accident which can be used to assess the radiation risk in real time and
provide helpful insight about its distribution.

Although the empirical Bayes estimation was applied to a single spatial
correlation parameter, the theory is more general to allow more parameters
to be estimated this way, e.g. a smoothness or a nugget parameter. For an
application of this approach to the isotropic spatial model see
\cite{roy16}. On the other hand, when many parameters are included, the
sampling and evaluation grids must be chosen carefully as a larger grid
takes longer to compute.

Another extension of our model, is the use of a spatially-varying temporal
autocorrelation parameter. Although this model allows us to capture the
influence across spatial components, it becomes challenging to fit as
there  are more parameters to estimate.

Our model for the latent process assumes linear transition in time.
Although not considered in this paper, it would be possible to apply the
methodology to non-linear models using local linear approximations. 

A potential research avenue is the application of this methodology to the
dynamic spatiotemporal design problem, see e.g. \cite{WiRo1} incorporating
parameter uncertainty in the design as well. Many interesting applications
can be found in the point-process framework and it would be interesting to
see how the suggested methodology performs in this case. Finally, the ideas
of this paper can be applied to other models beyond the spatial framework.


\section{Appendix}

\subsection{Proof of Lemma~\ref{prop:skewed}}

We examine the limits for $i = 1,\ldots,n$, of the ratio
\begin{equation*}
  \lim_{u \rightarrow \infty} \frac{p(\xbf_t =
    \mu_t + u\mathbf{e}_i|\xbf_{t-1},\ybf_t,\theta,\phi)} {p(\xbf_t =
    \mu_t - u\mathbf{e}_i|\xbf_{t-1},\ybf_t,\theta,\phi)} ,
\end{equation*}
where $\mu_t = G_t \beta + \alpha (\xbf_{t-1} - G_{t-1}
\beta)$ and $\mathbf{e}_i$ is a vector whose $i$th component is 1 and all
other components are 0. If the limit is 0 or $\infty$, then
the distribution is left or right skewed respectively.

Then, by the symmetry of the normal distribution around its mean,
\begin{align*}
  \lim_{u \rightarrow \infty} \frac{p(\xbf_t = \mu_t +
    u\mathbf{e}_i|\xbf_{t-1},\ybf_t,\theta,\phi)} {p(\xbf_t = \mu_t -
    u\mathbf{e}_i|\xbf_{t-1},\ybf_t,\theta,\phi)} & = \lim_{u
    \rightarrow \infty} \frac{p(\ybf_t|\xbf_t = \mu_t + u\mathbf{e}_i)
    p(\xbf_t = \mu_t + u\mathbf{e}_i|\xbf_{t-1},\theta,\phi)} {
    p(\ybf_t|\xbf_t = \mu_t - u\mathbf{e}_i) p(\xbf_t
    = \mu_t - u\mathbf{e}_i|\xbf_{t-1},\theta,\phi)} \\
  & = \lim_{u \rightarrow \infty} \frac{p(\ybf_t|\xbf_t = \mu_t +
    u\mathbf{e}_i)} { p(\ybf_t|\xbf_t = \mu_t - u\mathbf{e}_i)} \\
  & = \lim_{u \rightarrow \infty} \frac{p(y_{i,t}|x_{i,t} = \mu_{i,t} +
    u)} { p(y_{i,t}|x_{i,t} = \mu_{i,t} - u)} ,
\end{align*}
where the last limit is either 0 or $\infty$ since the distribution of
$y_{i,t}|x_{i,t}$ is skewed.

\subsection{Proof of Theorem~\ref{thm:est_phi_via_PF}}

The estimate of the sequential empirical Bayes factor,
$\hat B_{1:t}(\phi;\tilde\phi)$ in equation~\eqref{eq:37}, depends on the
associated sequential empirical Bayes factors,
$\mathbf{b}_t=(b_t^1,\ldots,b_t^K)$, on the coarse grid for
$\phi_k \in \Phi_K$. Consequently, we need to first establish the
convergence of $\hat b_t^k,\; k=1,\ldots,K$.

{
Because $\theta^{(k,l)}$ is drawn using Gibbs sampling, and because
$\xbf_t^{(k,l)}$ is sampled by importance sampling conditioned on
$\theta^{(k,l)}$, the sample $(\xbf_{0:t}^{(k,l)}, \theta^{(k,l)})$,
$l=1,\ldots,L_k$ is a Harris ergodic Markov chain for each
$k \in \{1,\ldots,K\}$ from the distribution
$p(\xbf_{0:t},\theta|\ybf_{1:t},\phi_k)$. 

Let $\lambda_k = L_k/ \sum L_{k'}$ and
$\Lambda_K = \{ \lambda_1, \ldots, \lambda_K \}$. Then the concatenated sample
$(x^{(k;l)}_{1:t}, \theta^{(k;l)})$,
$l=1,\ldots,L_k$, $k=1,\ldots,K$ is a Harris ergodic Markov chain from the
mixture distribution with components the
$p(\xbf_{0:t},\theta|\ybf_{1:t},\phi_k)$ and corresponding weights $\lambda_k$.
}
The probability that the
$(k,l)$th sample is drawn from the $k$th mixture component is given by
\begin{equation*}
  f(\xbf_{0:t}^{(k,l)},\theta^{(k,l)}|\ybf_{1:t}, \phi_k) = \frac{\lambda_k
    p(\xbf_{0:t}^{(k,l)},\theta^{(k,l)}|\ybf_{1:t},\phi_k)}
  {p_{\text{{mix}}}(\xbf_{0:t}^{(k,l)},\theta^{(k,l)}|\ybf_{1:t},  
    \Phi_K,\Lambda_K)},
\end{equation*}
where 
$p_{\text{mix}}(\xbf_{0:t},\theta|\ybf_{1:t},\Phi_K,\Lambda_K)$
denotes the mixture distribution of
$p(\xbf_{0:t},\theta|\ybf_{1:t},\phi_k)$ for $k = 1,\ldots,K$ with weights
$\lambda_k$ defined in~\eqref{eq:pmix}. Define the reverse logistic
log-likelihood 
\begin{equation} \label{eq:rev_log_reg}
  \ell (\mathbf{b}_t) = \sum_{k=1}^K \sum_{l=1}^{L_k} \log
  f(\xbf_{0:t}^{(k,l)},\theta^{(k,l)}|\ybf_{1:t}, \phi_k). 
\end{equation}
{Then using similar arguments as in \cite{BuDo} one may show that the maximizing argument of $\ell$, i.e. $\hat{\mathbf{b}}_t = \argmax  \ell(\mathbf{b}_t)$ converges a.s. to  the sequential empirical Bayes factors $\mathbf{b}_t$.
}

Next, observe that $\hat B_{1:t}(\phi;\tilde{\phi})$ can be written as
\begin{align} \label{eqn16} \sum_{k=1}^K \frac{1}{L_k} \sum_{l=1}^{L_k}
  \frac{ \lambda_k p(\xbf_{0:t}^{(k,l)}|\theta^{(k,l)},\phi)
  }{\sum_{k'=1}^K \frac{\lambda_{k'} }{\hat b^{k'}_t}
    p(\xbf_{0:t}^{(k,l)}|\theta^{(k,l)},\phi_{k'})}
  \stackrel{\mbox{a.s.}}{\longrightarrow} \sum_{k=1}^K \int \frac{
    \lambda_k p(\xbf_{0:t}|\theta,\phi) }{\sum_{k'=1}^K \frac{\lambda_{k'}
    }{b^{k'}_t} p(\xbf_{0:t}|\theta,\phi_{k'})}
  p(\xbf_{0:t},\theta|\ybf_{1:t},\phi_k) d(\xbf_{0:t},\theta).
    \end{align}
 The right hand side of equation~\eqref{eqn16}  equals 
 \begin{equation} \label{eqn17} B_{1:t}(\phi;\tilde \phi) \times
   \sum_{k=1}^K \int \frac{ \lambda_k p(\xbf_{0:t}|\theta,\phi)
     /p(\ybf_{1:t} | \phi) }{\sum_{k'=1}^K \lambda_{k'}
     p(\xbf_{0:t}|\theta,\phi_{k'})/p(\ybf_{1:t} | \phi_{k'})}
   p(\xbf_{0:t},\theta|\ybf_{1:t},\phi_k) d(\xbf_{0:t},\theta),
\end{equation}
and multiplying and dividing by $p(\ybf_{1:t}| \xbf_{0:t}) p(\theta)$, one
deduces that the finite sum of equation~\eqref{eqn17} equals 1. The proof
is thus complete.

\section*{References}
\bibliographystyle{apalike}
\renewcommand{\thebtauxfile}{sto_final_main}
\begin{btSect}{}
\btPrintCited
\end{btSect}

\clearpage
\setcounter{section}{0}
\setcounter{figure}{0}
\setcounter{page}{1}

\chead{\scshape Web-based Supplementary Materials}

\begin{center}
  \LARGE
  Sequential Empirical Bayes Method for Filtering Dynamic
  Spatiotemporal Processes \\ Web-based Supplementary Materials\\[20pt]
  \Large
  Evangelos Evangelou$^{1}$
  \hspace{1pt}
  and Vasileios Maroulas$^{2}$\\[5pt]
  \normalsize
$^1$ {Department of Mathematical Sciences, University of Bath, Bath BA2 7AY, UK.}\\
$^2$ {Department of Mathematics, University of Tennessee, Knoxville, TN 37996, USA.}
\end{center}
\bigskip

\begin{center}
  \bfseries  
  \Large Simulation Results
\end{center}
\bigskip

The general setup of our simulations is as follows. The spatial dimension
is the closed interval $[0,1]$ and the spatial sampling locations consist
of $n=11$ equidistant points covering the spatial domain. The final
sampling time is denoted by $T$. The latent spatiotemporal process $\xbf_t$
is simulated with constant mean $\beta = 1$, autoregressive coefficient
$\alpha = 0.5$, and variance $\sigma^2 = 1$. The correlation between
components of $\xbf_t$ is calculated using the exponential spatial
correlation function, i.e.
\begin{equation*}
  \mathrm{Corr}(\xbf_{i,t},\xbf_{j,t}) = \exp(-d_{ij}/\phi),
\end{equation*}
where $d_{ij}$ stands for the distance between the $i$th and $j$th grid
point and $\phi = 0.4$ is the range parameter. At each time $t$ we simulate
a response $\ybf_t$ conditioned on the simulated $\xbf_t$ such that
$y_{i,t} \sim \mathrm{Poisson}(\tau e^{x_{i,t}})$ independently for each
$i$, for given $\tau$.

For inference, the priors specified in~\eqref{eq:2} were used with
$a_0 = 0$, $s_0 = 0.1$, $b_0 = 0$, $q_0 = 0.01$, $c_0 = 3$, and
$r_0 = 1/3$. The fine grid $\Phi$ consisted of $J=41$ equidistant points
between $\phi/2$ and $2\phi$, i.e.
$\Phi = \{0.200, 0.215, 0.230, \ldots, 0.800\}$ and the coarse grid to
$\Phi_K = \{0.230, 0.335, 0.440, 0.545, 0.650, 0.755\}$. The first element
of $\Phi_K$ corresponds to $\tilde{\phi}$.

Algorithm~\ref{alg:pf_phi} was run with Monte-Carlo sizes $L_k = L$ for
$k=1,\ldots,K$ and Algorithm~\ref{alg:gibbs1} with Gibbs iterations $L_{g}$
and particle size $N$.

\section{Effect of the proposal distribution}
\label{sec:2}

In this section we compare the three choices of the proposal distribution
discussed in the paper: (a)~the Gaussian proposal; (b)~the copula mean-only
skewness correction; and (c)~the copula mean-and-skewness
correction.

The time dimension was $T=100$. We performed 30 simulations from the model
with $\tau=1$.
This model choice ensures that there is a substantial amount of skewness in
the observations and will make the comparison between the three proposals
more apparent. We measure the skewness of the approximation by computing
the parameter $\delta^2 = a^2/(1+a^2)$ such that values of $\delta^2$ close
to $1$ give large skewness and values close to $0$ give low skewness. In
our simulations, the skew-normal parameter $\delta^2$ had an average value
of 0.12 with the largest value being about 0.65.

The Monte-Carlo sizes were $L=100$, $N=100$,
and $L_\mathrm{g}=50$.

\begin{figure}
  \centering
  \includegraphics[width=.6\linewidth]{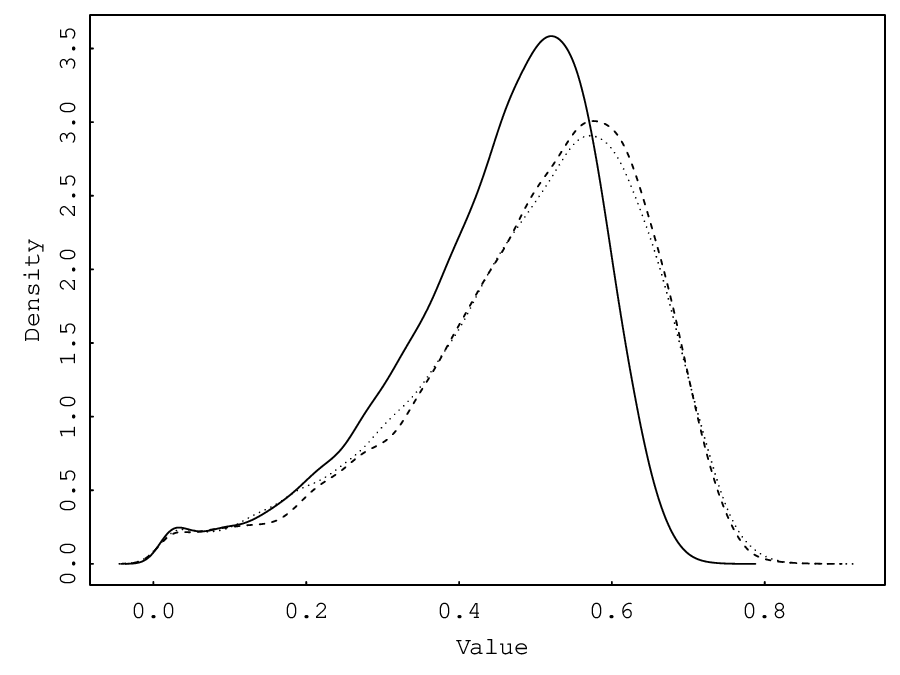}
  \caption{Density plots of the average ESS divided by the total number of
    generaged particles for: Gausian proposal (solid); Gaussian with
    mean-only correction (dashed); copula with mean-and-skewness correction
    (dotted).}
  \label{fig:ESS_plot}
\end{figure}

For each time iteration we compute the effective sample size (ESS) for each
method. Ideally we want $\mathrm{ESS}$ to be close to $N$ which will
indicate that the proposal distribution generates good samples while a very
low ESS would indicate degeneracy in the particles, which is not uncommon
in high dimensions. Figure~\ref{fig:ESS_plot} shows a density plot for the
distribution of the average ESS over the $L$ samples at each time iteration
and for the 30 simulations (i.e, $30 \times T$ values), expressed as a
proportion of the total number of samples $N$ for each of the three
proposal distributions. As shown in the figure, the uncorrected Gaussian
proposal has a significantly lower ESS that the two corrected methods but
the two skewness correction methods are very similar. 
Based on our results, and in the following, we consider the
mean-only corrected proposal only.

\section{Comparison with the simplified Bayes factor estimator}
\label{sec:gridphi}

The simplified Bayes factor estimator is given in~\eqref{eq:16}. This
estimator simulates conditioned on $\phi = \tilde{\phi}$ only and uses
these samples to compute the Bayes factor estimate for all $\phi \in \Phi$.
In this case the reverse logistic estimates are not needed. However, as we
discuss in Remark~\ref{naMC}, this can potentially introduce bias if the
true $\phi$ is far from $\tilde{\phi}$.

\begin{figure}
  \centering
  \includegraphics[width=1\linewidth]{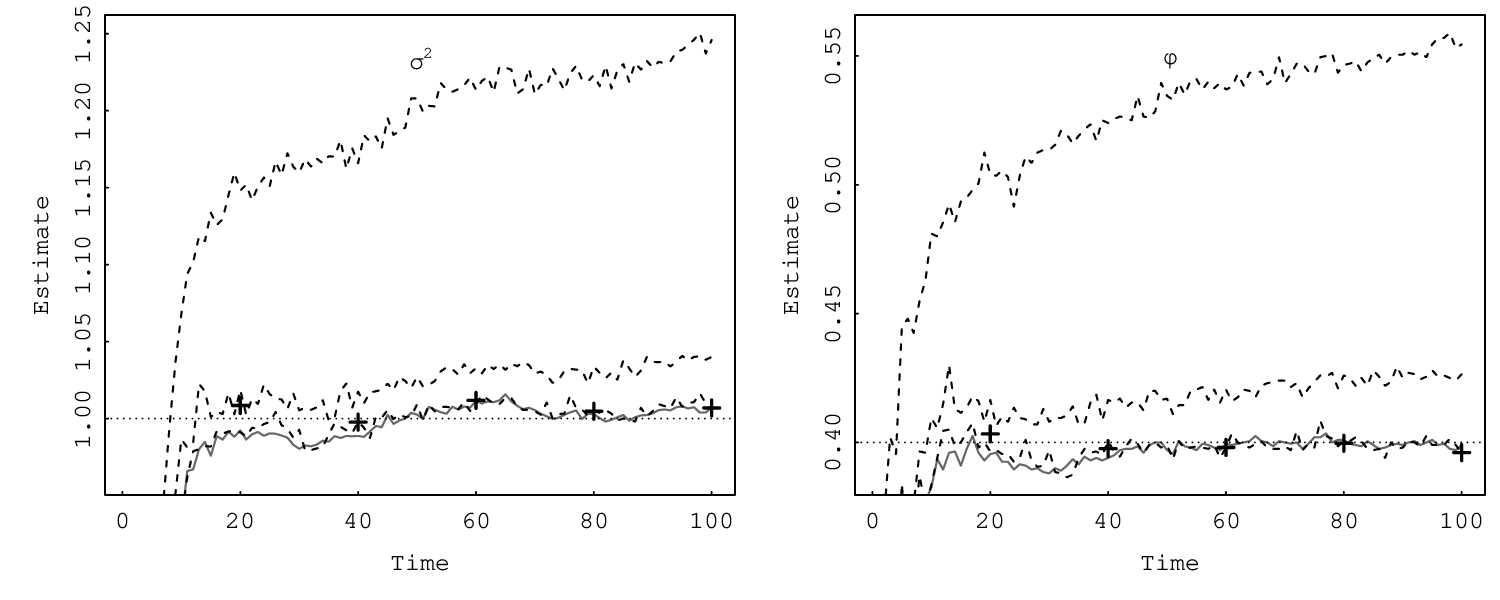}
  \caption{Average estimates for the parameters $\sigma^2$ and $\phi$ from
    the different Bayes factor estimators. The solid line is the mixed
    Bayes factor estimator~\eqref{eq:37}. The dashed lines correspond to
    the simplified Bayes factor estimator~\eqref{eq:16} for
    $\tilde{\phi} = 0.395, 0.500, 0.710$ from bottom to top. The average
    offline estimates are shown by $+$ and the true parameter value is
    shown by a horizontal line.}
  \label{fig:snmc_simpl}
\end{figure}

In this section we compare the bias of the simplified Bayes factor
estimator with the proposed estimator~\eqref{eq:37} for the same Poisson
model used in Section~\ref{sec:2} but with increased $L=500$ and $\tau=10$. We
consider~\eqref{eq:16} with three different values of
$\tilde{\phi} = 0.395, 0.500, 0.710$, where the first value is very close
to the true $\phi$, the second value is at the middle of the range of
$\Phi$, and the third value is  far from the true.
The
simplified Bayes factor estimator was tested on  30 simulated
cases  with burn-in
$L_\text{bi} = 50$, thinning $L_\text{th} = 10$ and final sample size 3000. The
average estimate over the 30 cases for each method was computed for each time
point. This is plotted in Figure~\ref{fig:snmc_simpl} for the parameters
$\sigma^2$ and $\phi$. The estimation for the parameters $\alpha$ and
$\beta$ did not show any obvious discrepancy.

Our results verify that the simplified Bayes factor estimator is biased and
this is more apparent when $\tilde{\phi}$ is far from the true $\phi$.
Although the estimation for $\phi$ and $\sigma^2$ is biased, this does not
seem to influence the estimation of $\alpha$ and $\beta$. This phenomenon
has been observed elsewhere in the literature for the spatial-only case
\citep[see][]{zhang2002estimation}. Based on our results, the mixed Bayes
factor estimator is recommended instead of the simplified one.

\section{Estimation performance}
\label{sec:3}

In this section we assess the estimation performance of the proposed
algorithm with the mean-only corrected proposal. We use the same setting as
in Section~\ref{sec:gridphi} and the same 30 simulated data. We compare
our estimates against an offline MCMC algorithm with the same Monte-Carlo sizes as
Section~\ref{sec:gridphi}.
The offline MCMC algorithm uses Gibbs sampling for the parameters $\theta$
and a Metropolis-Hastings step to update $\phi$ and $x_{i,t}$, the $i$th
component of $\xbf_t$ conditioned on everything else. The prior for $\phi$
was the exponential distribution with mean 0.4 and the Metropolis-Hastings
step was selected for acceptance between 0.2 to 0.4. Convergence
diagnostics of the MCMC output did not indicate any issues.

Because of the increasing computational time, we only ran the offline
algorithm for selected time points $T_i = 20, 40, 60, 80, 100$, where at
each time only data up to $T_i$ were observed to make the results
comparable with the online method.

\begin{figure}
  \centering
  \includegraphics[width=1\linewidth]{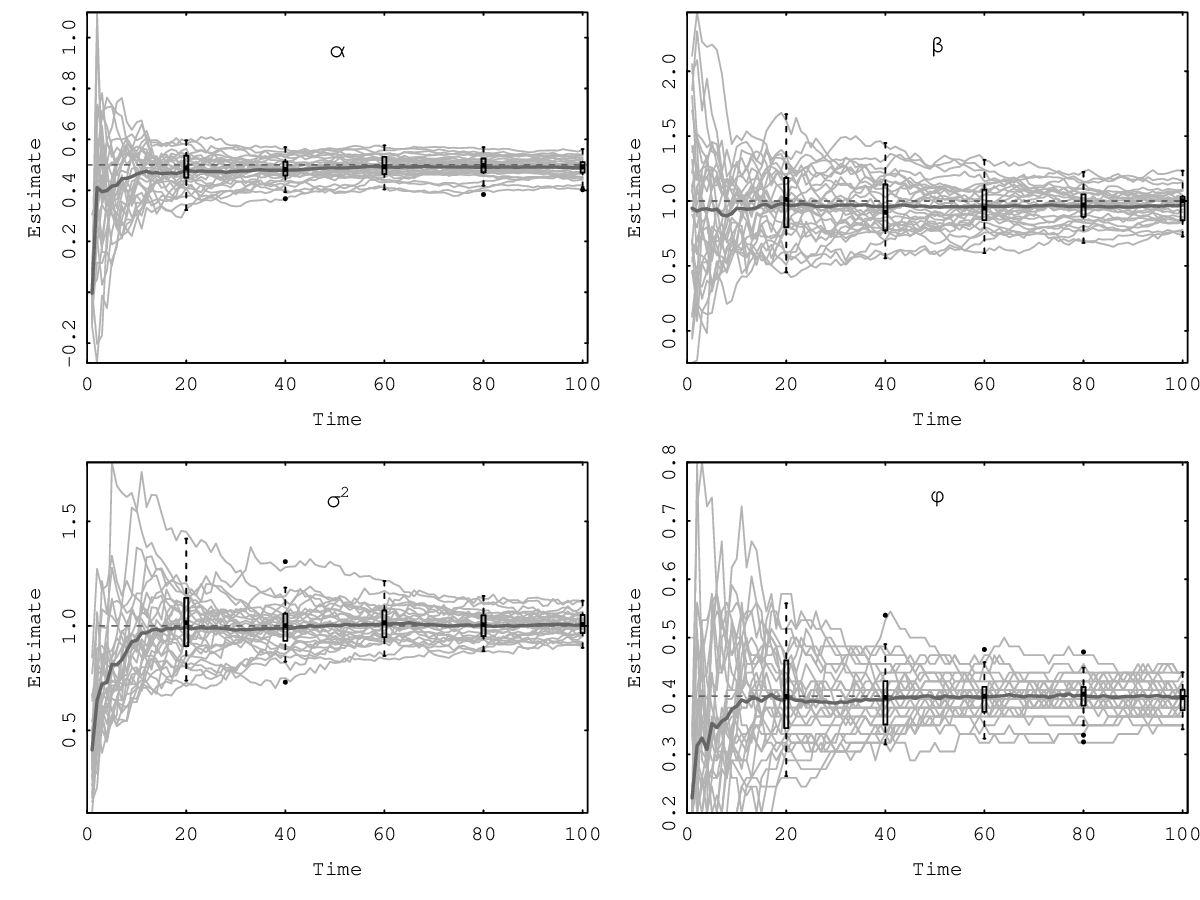}
  \caption{Comparison of estimates between the proposed online algorithm
    and the offline MCMC algorithm for the four parameters $\alpha$,
    $\beta$, $\sigma^2$, $\phi$. The light gray lines show the estimate for
    every simulation and the dark gray is the mean over all simulations.
    The boxplots show the distribution of the estimates using an offline
    MCMC algorithm with data available up to that time. The true parameter
    value is shown by a dashed line.}
  \label{fig:snmc_est}
\end{figure}

In Figure~\ref{fig:snmc_est} we plot the parameter estimates for each
parameter in time for the online algorithm for each simulation and the
distribution of the offline estimates from all simulations at the selected
time points. It can be seen that the distributions from the two methods are
very similar. In particular, the variability of our estimates reduces as
we see more data and the bias is reduced which is a desirable property.

\begin{figure}
  \centering
  \includegraphics[width=.6\linewidth]{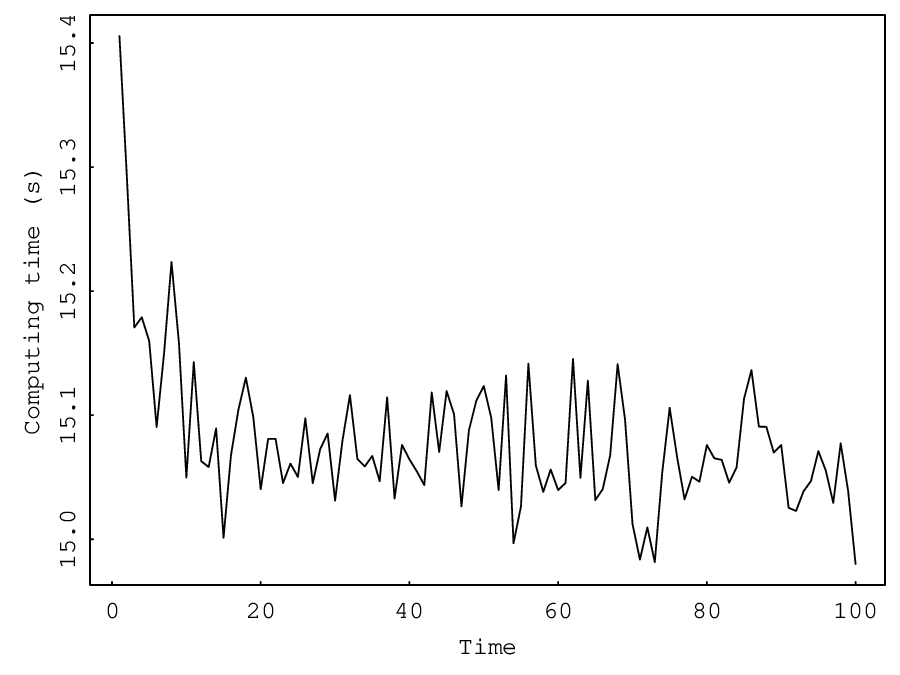}
  \caption{Average computing time per time iteration for the sequential
    empirical Bayes algorithm across simulations. The computations were
    performed on computer with Intel Core i5-2500 3.30GHz CPU and 4Gb RAM.}
  \label{fig:snmc_calctime}
\end{figure}

For each time iteration, the computing time for the sequential empirical
Bayes algorithm was recorded, i.e. one iteration of
Algorithm~\ref{alg:pf_phi}, and the average over the 30 simulations was
taken. The average computing time is shown in Figure~\ref{fig:snmc_calctime}. It
can be seen that the computing time does not increase in time as one would
expect from an online algorithm. 

\begin{figure}[!htb]
  \centering
  \includegraphics[width=\linewidth]{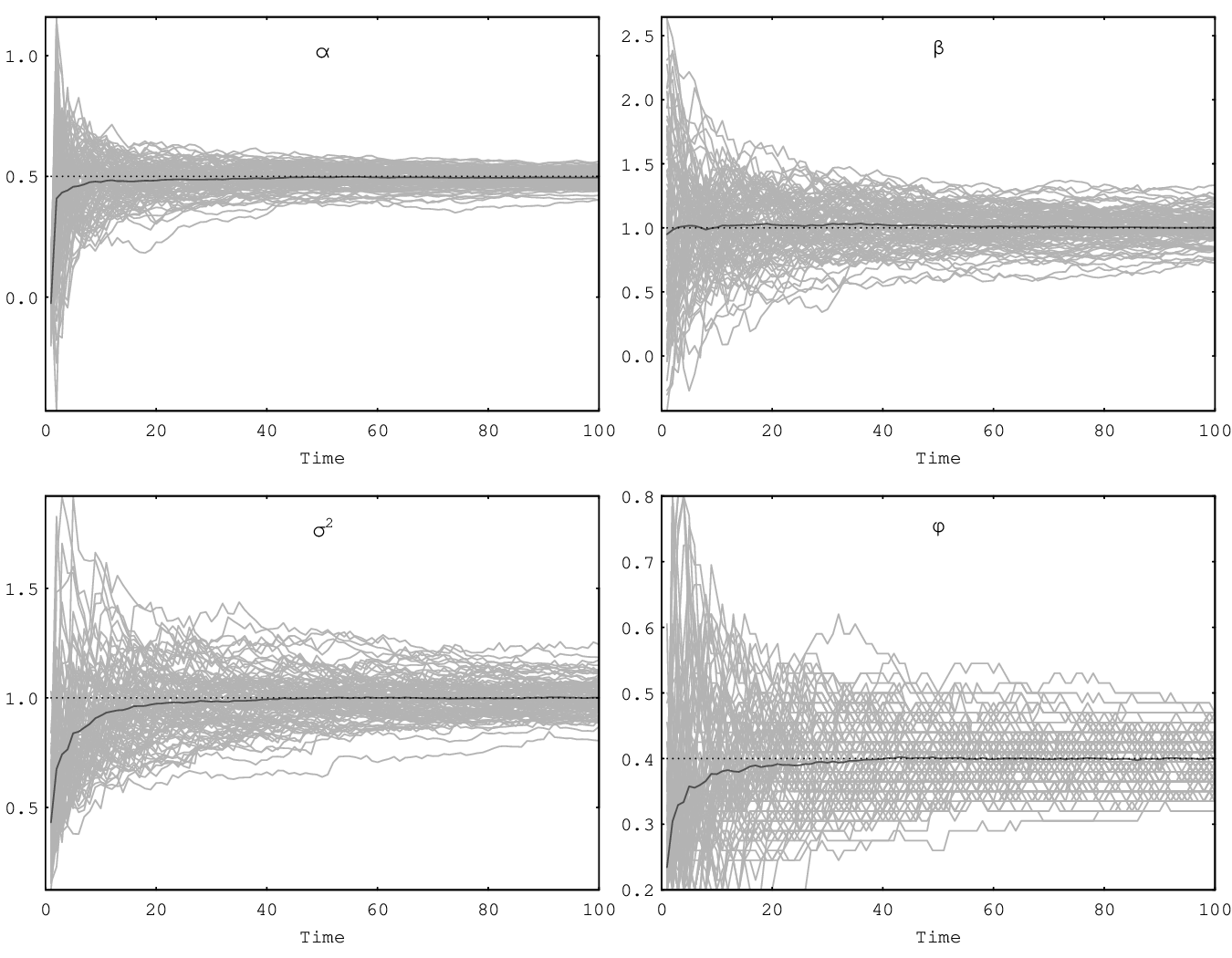}
  \caption{Estimates of $\alpha$, $\beta$, $\sigma^2$ and
    $\phi$ in that order across time. The dotted line shows the true
    parameter value, and the light gray lines represent the estimates
    corresponding to each simulation across time. The mean across all
    simulations is shown by a dark gray line.}
  \label{fig:2}
\end{figure}

\begin{figure}[!htb]
  \centering
  \includegraphics[width=\linewidth]{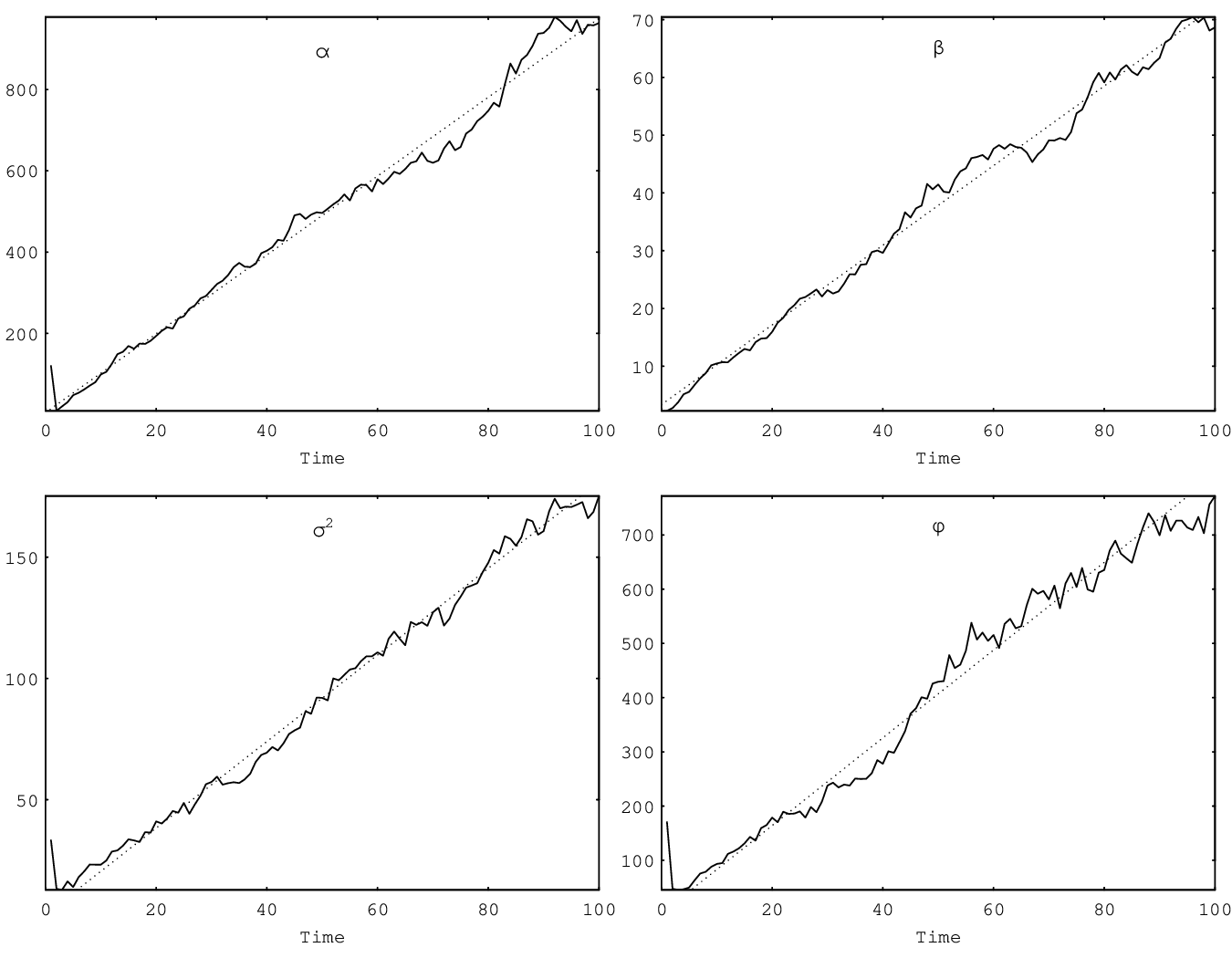}
  \caption{Reciprocal of the sample variances across simulations of the
     estimates for $\alpha$, $\beta$, $\sigma^2$ and $\phi$ in
    that order across time. The dotted line indicates the linear rate
    increase in time for the reciprocal of the sample variance.}
  \label{fig:2var}
\end{figure}

Subsequently, the number of simulations was increased to 100, but in this
case only the online algorithm was computed. This was to assess any
potential bias in our method. The results from these simulations are shown
in Figure~\ref{fig:2} which show no apparent bias. On average, across all
simulations, the four parameters are estimated accurately. Note the
convergence of the estimates towards the true value and the reduction of
uncertainty as more data are observed which demonstrates the suitability of
our method. The sample variance of our estimates across simulations at each
time point was calculated and its reciprocal was plotted against time. The
plots  corresponding to the four parameters are shown in
Figure~\ref{fig:2var}. It can be seen that the variability decreases
linearly which indicates a reduction in the length of the posterior
credible interval in the order of $1/\sqrt{t}$ as time $t$ increases.

\section{Simulation with longer time span}
\label{sec:4}

In this section we use simulated data to compare the proposed algorithm
against an offline MCMC algorithm. In this example the data were simulated
from the model of Section~\ref{sec:gridphi} but with final time increased to
$T=1000$. Only one sample was generated in this example.
The data were subsequently fitted using the proposed online algorithm and
an offline MCMC {smoothing} algorithm. The priors for both methods were the
same as in Section~\ref{sec:2} and a Monte Carlo sizes were as in
Section~\ref{sec:gridphi}. 

\begin{figure}[h]
  \centering
  \includegraphics[height=.4\textheight]{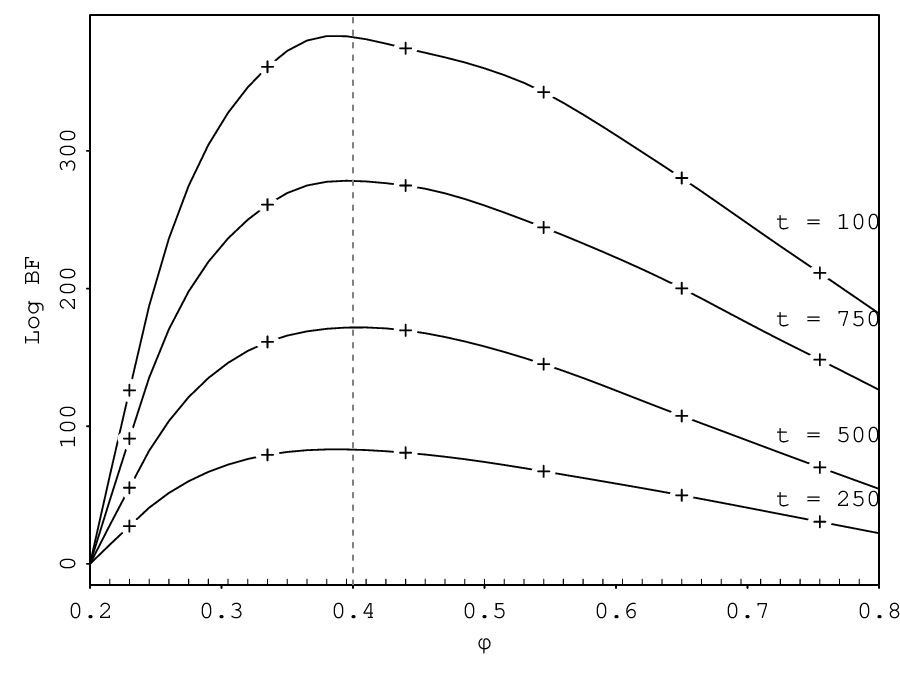}
  \caption{Logarithm of the Bayes factor, $\log B_t(\phi;\tilde \phi)$,
    plotted against $\phi$ for different times. The true $\phi$ is shown by
    a vertical line. The coarse grid $\Phi_K$ is marked by $+$ on each line
    and the lines at the bottom show the fine grid $\Phi$.}
  \label{fig:exlogb}
\end{figure}

Figure~\ref{fig:exlogb} shows the function
$\log B_{1:t}(\phi; \tilde{\phi})$ computed by the proposed online
algorithm for selected values of $t$, {along with the grids
  $\Phi_K$ and $\Phi$}. The maximizer of this function is the estimate for
$\phi$ at time $t$. Note that, as $t$ increases, the maximum of this
function converges to the true value and the uncertainty is reduced. To
derive a confidence interval we view $B_{1:t}(\phi; \tilde{\phi})$ as an
unnormalized posterior pdf for $\phi$ and the corresponding cumulative sum
is the unnormalized cumulative distribution function (cdf). We then
approximate the corresponding quantiles by polynomial interpolation of
$\phi$ against the normalized cdf.

\begin{figure}[h]
  \centering
  \includegraphics[width=\linewidth]{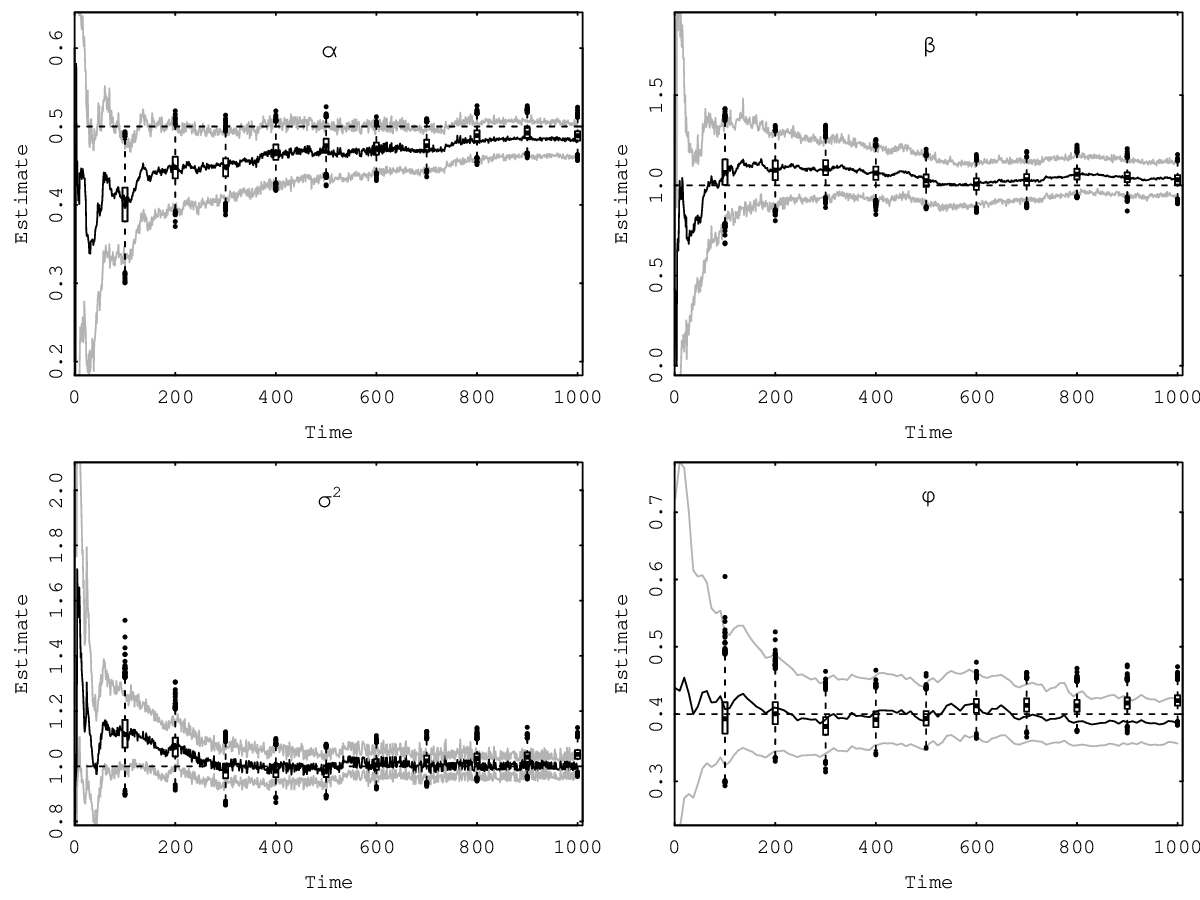}
  \caption{Parameter estimates across time using the proposed online
    algorithm (black lines) with 99\% confidence intervals (light grey
    lines), and an offline algorithm (boxplots). The true parameter value
    is shown by a horizontal line.}
  \label{fig:exest}
\end{figure}

The estimates (MC average for $\theta$, EB estimate for $\phi$) and 99\%
credible intervals (MC quantiles for $\theta$, polynomial interpolation for
$\phi$) for each parameter using data $\ybf_{1:t}$ across $t$ are plotted
in Figure~\ref{fig:exest}. As shown in the figure, since both algorithms sample
from the same posterior distribution, conditioned on $\ybf_{1:t}$, the
estimates and credible intervals obtained between them are very similar and
capture the true parameter values even for a longer time span. 

\section*{References}
\bibliographystyle{apalike}
\renewcommand{\thebtauxfile}{sto_final_suppl}
\begin{btSect}{}
\btPrintCited
\end{btSect}

\end{document}